\documentclass[floats,floatfix,showpacs,amssymb,prd,twocolumn,superscriptaddress,nofootinbib]{revtex4-1}

\newlength{\figw} 
\usepackage{graphicx,epsf, epsfig, amssymb}
\usepackage{bm}
\usepackage{longtable,tabularx}
\usepackage{color}
\usepackage[breaklinks]{hyperref}
\usepackage{amsfonts,amsmath,amssymb,mathrsfs,gensymb,mathtools}
\usepackage{natbib}
\usepackage{aas_macros}
\usepackage{array}
\usepackage{multirow,enumitem}
\usepackage{rotating,array}
\usepackage{graphicx}
\usepackage{capt-of}
\usepackage{makecell}
\usepackage[usenames,dvipsnames]{xcolor}

\DeclareMathOperator\sign{sgn}
\newcommand\optional[1]{}

\begin{document}

\title{Multi-timescale analysis of phase transitions in precessing black-hole binaries}

\author{Davide Gerosa}
\email{d.gerosa@damtp.cam.ac.uk}
\affiliation{Department of Applied Mathematics and Theoretical Physics, Centre for Mathematical Sciences, University of Cambridge, Wilberforce Road, Cambridge CB3 0WA, UK}

\author{Michael Kesden}
\email{kesden@utdallas.edu}
\affiliation{Department of Physics, The University of Texas at Dallas, Richardson, TX 75080, USA }

\author{Ulrich Sperhake}
\email{u.sperhake@damtp.cam.ac.uk}
\affiliation{Department of Applied Mathematics and Theoretical Physics, Centre for Mathematical Sciences, University of Cambridge, Wilberforce Road, Cambridge CB3 0WA, UK}
\affiliation{Department of Physics and Astronomy, The University of
Mississippi, University, MS 38677, USA}
\affiliation{California Institute of Technology, Pasadena, CA 91125, USA}

\author{Emanuele Berti}
\email{eberti@olemiss.edu}
\affiliation{Department of Physics and Astronomy, The University of
Mississippi, University, MS 38677, USA}
\affiliation{CENTRA, Departamento de F\'isica, Instituto Superior
T\'ecnico, Universidade de Lisboa, Avenida Rovisco Pais 1,
1049 Lisboa, Portugal}

\author{Richard O'Shaughnessy}
\email{rossma@rit.edu}
\affiliation{Center for Computational Relativity and Gravitation, Rochester Institute of Technology, Rochester, NY 14623, USA}


\date{\today}

\begin{abstract}
The dynamics of precessing binary black holes (BBHs) in the post-Newtonian regime has a strong timescale hierarchy: the
orbital timescale is very short compared to the spin-precession timescale which, in turn, is much shorter than the
radiation-reaction timescale on which the orbit is shrinking due to gravitational-wave emission.  We exploit this timescale
hierarchy to develop a multi-scale analysis of BBH dynamics elaborating on the analysis of Kesden {\it et al.} \cite{2015PhRvL.114h1103K}.
We solve the spin-precession equations analytically on the precession time and then implement a quasi-adiabatic
approach to evolve these solutions on the longer radiation-reaction time. This procedure leads to an innovative
``precession-averaged'' post-Newtonian approach to studying precessing BBHs.  We use our new solutions to classify
BBH spin precession into three distinct morphologies, then investigate phase transitions between these morphologies as
BBHs inspiral.  These precession-averaged post-Newtonian inspirals can be efficiently calculated from arbitrarily large
separations, thus making progress towards bridging the gap between astrophysics and numerical relativity. 
\end{abstract}
\maketitle

\section{Introduction}
\label{sec:intro}

Observations suggest that astrophysical black holes are generally spinning \cite{2013CQGra..30x4004R,
2011CQGra..28k4009M,2015PhR...548....1M} and can form binary systems \cite{2015ASSP...40..103B}.  Spinning binary
black holes (BBHs) are a promising source of gravitational waves (GWs) \cite{2010CQGra..27q3001A,
2015JPhCS.610a2001B,2015ASSP...40..147S} for current and future detectors \cite{2010CQGra..27h4006H,
2013IJMPD..2241010U, 2012CQGra..29l4007S, 2010CQGra..27s4002P, 2013CQGra..30v4010M, 2013PASA...30...17M, 
2009arXiv0909.1058J, 2013CQGra..30v4009K}.   BBH dynamics
is remarkably complex and interesting, especially when both BBHs are spinning.  BBH systems have three angular
momenta, the two spins and the orbital angular momentum, all coupled to each other.  Spin-orbit and spin-spin couplings
cause these angular momenta to precess, changing their orientation in space on the precession timescale
\cite{1994PhRvD..49.6274A,1995PhRvD..52..821K}.  On the longer radiation-reaction timescale, GWs take energy and
momentum out of the system, thus shrinking the orbit \cite{1963PhRv..131..435P, 1964PhRv..136.1224P}.  These emitted
GWs encode all the richness of the precessional dynamics but are also more challenging to detect and characterize than
GWs emitted by non-precessing systems
\cite{2012PhRvD..86f4020B,2014PhRvD..89b4010H,2015PhRvD..91f2010D,2015PhRvD..91d2003V,2014PhRvD..90b4018F,2014PhRvD..89j2005O,2014PhRvD..89j4023C,2015ApJ...798L..17C}.  

Expanding on the analysis in our previous paper  \cite{2015PhRvL.114h1103K}, we introduce a multi-timescale analysis of the dynamics of spinning, precessing BBHs.  Multi-timescale
analyses are commonly used in binary dynamics.  For example, in eccentric binaries the orbital period, periastron
precession, and radiation-reaction timescales usually differ by orders of magnitude; the dynamics of these systems can be
studied using techniques that explicitly exploit this timescale hierarchy, such as osculating orbital elements
\cite{1990PhRvD..42.1123L} or the variation of constants \cite{2004PhRvD..70f4028D}.  Exploiting timescale hierarchies
leads to deeper understanding of the dynamics because different physical processes are decoupled and individually
analyzed.  

Precessing BBHs evolve on three distinct timescales:
\begin{enumerate}

\item BBHs orbit each other (changing the binary separation $\mathbf{r}$) on the orbital time 
$t_{\rm orb}\sim r^{3/2} /(GM)^{1/2}$,

\item the spins and the orbital angular momentum change direction on the precession time
\mbox{$t_{\rm pre} \sim c^2 r^{5/2}/(GM)^{3/2}$},

\item the magnitudes of the orbital energy and angular momentum decrease on the radiation-reaction time
$t_{\rm RR} \sim c^5 r^{4} /(GM)^3$.

\end{enumerate}
Here $r = |\mathbf{r}|$ is the magnitude of the binary separation, 
 $M$ is the 
total mass of the binary, and prefactors of order unity have been omitted.  In the post-Newtonian (PN) regime, 
$r \gg GM/c^2$  and these timescales are widely separated:
\begin{align}
\label{hierarchy}
t_{\rm orb} \ll t_{\rm pre} \ll t_{\rm RR}\,.
\end{align}
BBHs complete many orbits before their angular momenta appreciably precess, and the angular momenta complete many
precession cycles before the separation decreases significantly.

The first inequality ($t_{\rm orb} \ll t_{\rm pre}$) has been widely exploited to understand spin dynamics and approximate the
GW signal.  This approximation forms the foundation of the orbit-averaged spin-precession equations for adiabatic
quasicircular orbits examined extensively in the pioneering study of Apostolatos et al.~\cite{1994PhRvD..49.6274A} and later extended by Arun et
al.~\cite{2009PhRvD..79j4023A,2011PhRvD..84d9901A}.
Using these equations, several authors have systematically explored the physics of spin precession and their implications
for GW detection \cite{2006PhRvD..74l2001L,2011PhRvD..84b2002L} and astrophysics \cite{2012PhRvD..85l4049B,
2010PhRvD..81h4054K}.  Following the early work by Schnittman on spin-orbit resonances \cite{2004PhRvD..70l4020S},
PN spin dynamics has been used to predict the final spins \cite{2010PhRvD..81h4054K} and recoils
\cite{2010ApJ...715.1006K,2012PhRvD..85l4049B} following BBH mergers,  select initial conditions for numerical-relativity
simulations \cite{2015PhRvL.114n1101L}, characterize formation scenarios for stellar-mass BH binaries
\cite{2013PhRvD..87j4028G}, and address the distinguishability of these scenarios by future GW observations
\cite{2014PhRvD..89l4025G, 2014PhRvL.112y1101V,2014CQGra..31j5017G}.

The second inequality ($t_{\rm pre}\ll t_{\rm RR}$) has received less attention because until now there were no explicit
solutions for generic spin precession (unlike the Keplerian orbits that readily allowed orbit averaging in previous work). 
Our new solutions for spin precession allow us to fully exploit the timescale hierarchy of Eq.~(\ref{hierarchy}), 
expanding and detailing the ideas put forward in our previous \emph{Letter} \cite{2015PhRvL.114h1103K}.  We showed
that spin precession is quasi-periodic implying that the relative orientations of the three angular
momenta are fully specified by a single parameter, the magnitude of the total spin, that oscillates on the precession time.
As is common in multi-timescale analyses, once the dynamics on the shorter time has been solved, the behavior of the
system on the longer time scale can be studied as a quasi-adiabatic process.  We evolve our precessional solutions during
the inspiral by double-averaging the PN equations over both the orbital and the precessional timescales. Semi-analytical
precession-averaged inspirals turn out to be extremely efficient and can be carried out from infinitely large separation
with negligible computational cost.

While our focus in this work is on spin precession, our study benefits from several recent investigations which also used
separation of timescales to efficiently and accurately approximate both the dynamics and the associated GW signal. 
A series of papers by Klein et al.~\cite{2013PhRvD..88l4015K,2014PhRvD..89j4023C,2014PhRvD..90l4029K}
used a multi-scale analysis to construct semianalytic approximations to the frequency-domain waveforms for generic
two-spin precessing binaries.  Lundgren and O'Shaughnessy \cite{2014PhRvD..89d4021L} used this timescale hierarchy
to construct semianalytic approximations to the inspiral of precessing binaries with a single significant spin.
The GWs emitted during the full inspiral-merger-ringdown of spinning,
  precessing binaries were also investigated using phenomenological
models based on a single ``effective spin'' approximation~\cite{2012PhRvD..86j4063S,2014PhRvL.113o1101H,2015PhRvD..91b4043S}
and the effective-one-body framework~\cite{2014PhRvD..89h4006P}.

This paper is organized as follows. In Section~\ref{sec:prectime} we derive explicit solutions for generic BBH spin
precession at 2PN order on timescales short compared to the radiation-reaction time $t_{\rm RR}$.  These solutions allow spin
precession to be classified into three different morphologies characterized by the qualitative behavior of the angle between
the components of the two spins in the orbital plane.  In Sec.~\ref{sec:insptime}, we use our new solutions to precession
average the radiation reaction on the binary at 1PN order and demonstrate how this precession averaging improves 
the computational efficiency with which GW-induced inspirals can be calculated compared to previous approaches relying 
solely on orbit averaging. Precession-averaged evolution does not preserve the memory of the initial precessional
phase, just like orbit-averaged PN evolutions do not track the orbital phase. Sec.~\ref{sec:phtr} explores phase transitions between the three precessional morphologies, 
which are readily identified using our new formalism and have potentially interesting observational consequences.  Finally, we
conclude in Sec.~\ref{sec:concl}, highlighting the relevance of our new PN approach to both the theoretical understanding 
of BBHs and observational GW astronomy. We mainly focus on the relative orientation of the momenta; 
the evolution of the global orientation of the system will be addressed elsewhere \cite{ZhaoPrep}. 
Throughout the paper, we use geometrical units ($G=c=1$) and write vectors in
boldface, denoting the corresponding unit vectors by hats and their magnitude as (e.g.)  $L=|\mathbf{L}|$. Latin subscripts ($i=1,2$) label the BHs in
the binary. Binaries are studied at separations $r\geq 10M$, taken as a simple but {\it ad hoc}  threshold for the breakdown of the  PN approximation \cite{2006PhRvD..74j4005B,2009PhRvD..79h4010C,2009PhRvD..80h4043B}. Animated versions of some figures are available online at the URLs listed in Ref.~\cite{DGwebsite}.

\section{Analytic solutions on the precessional time scale}
\label{sec:prectime}

In this section we focus on the binary dynamics on the precessional time. Angular momentum conservation
(Sec.~\ref{Jconserve}) and the existence of a further constant of motion   (Sec.~\ref{subseceffpot}) provide a simple
parametrization of the binary dynamics through the identification of effective potentials. Solutions are classified according to
the precession geometry (Sec.~\ref{subsecmorph}) and eventually expressed in an inertial frame
(Sec.~\ref{timedependent}).

\subsection{Parametrization of precessional dynamics}
\label{Jconserve}

Let us consider BBHs on a circular orbit.\footnote{Our approach can be readily generalized to nonzero eccentricity
without complicating the geometry since the orbital pericenter precesses on a shorter timescale than the BBH spins do. 
We restrict our attention to circular orbits since radiation reaction is expected to suppress the eccentricity at large
separations for most astrophysical systems  \cite{1963PhRv..131..435P, 1964PhRv..136.1224P}.}  Let $m_1$ and $m_2$
denote the BBH masses, in terms of which we can define the mass ratio $q=m_2/m_1\leq 1$, the total mass
$M=m_1+m_2$, and the symmetric mass ratio $\eta=m_1 m_2 / M^2$. The spin magnitudes $S_i=m_i^2 \chi_i$ ($i=1,2$)
are most conveniently parametrized in terms of the dimensionless Kerr parameter $0\leq\chi_i\leq1$, while the magnitude
of the orbital angular momentum is related to the binary separation $r$ through the Newtonian expression
$L=\eta (r M^3)^{1/2}$.

\begin{figure}
\includegraphics[width=\columnwidth]{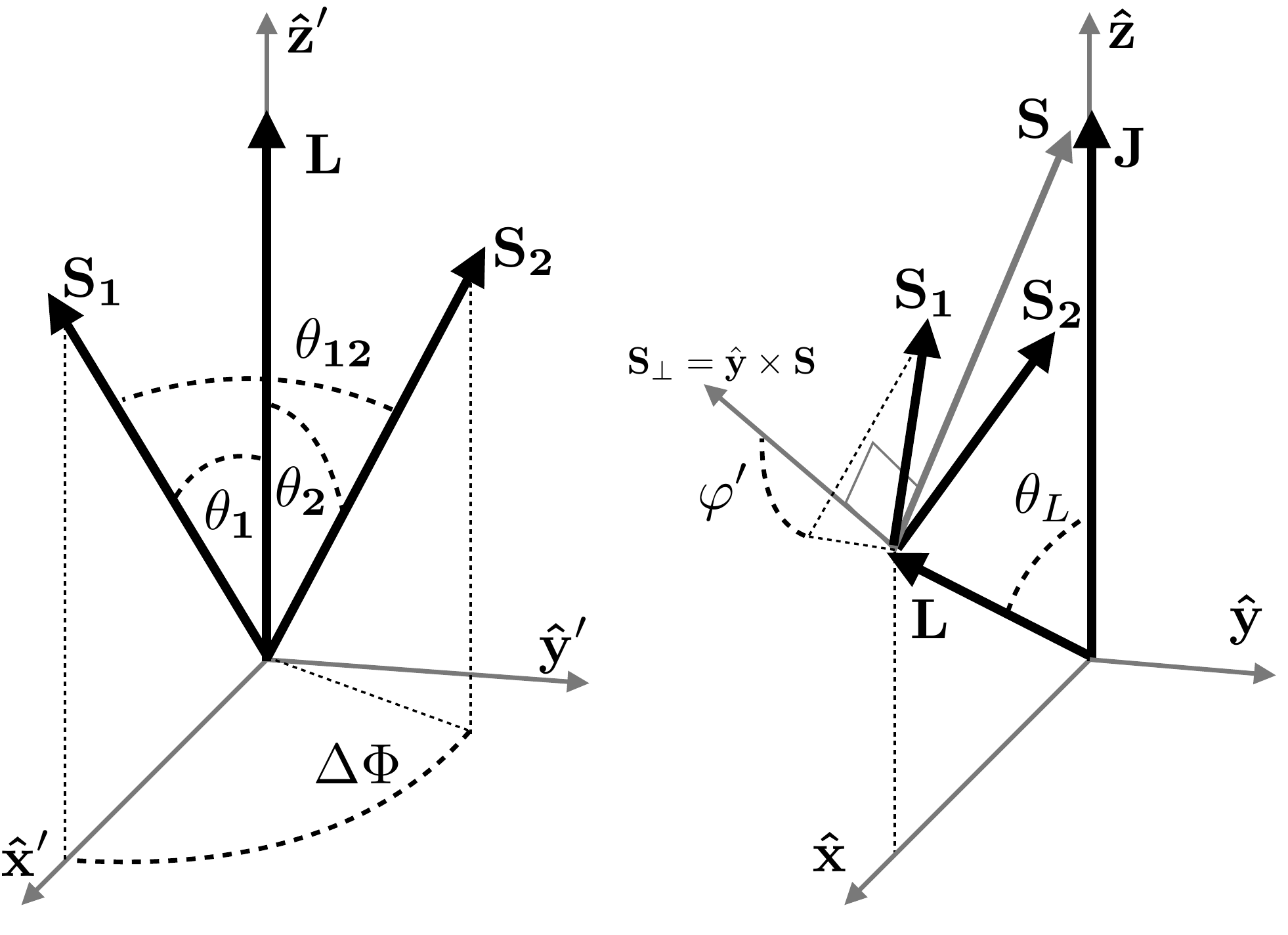}
\caption{Reference frames used in this paper to study BBH spin precession. The angles $\theta_1$, $\theta_2$,
$\Delta\Phi$, and $\theta_{12}$ are defined in a frame aligned with the orbital angular momentum $\mathbf{L}$ (left panel).
The binary dynamics can also be studied in a frame aligned with the total angular momentum $\mathbf{J}$ (right panel). 
Once $\mathbf{L}$ is taken to lie in the $xz$-plane, its direction is specified by $S$ through the angle $\theta_L$.  The
angle $\varphi'$ corresponds to rotations of $\mathbf{S_1}$ and $\mathbf{S_2}$ about the total spin $\mathbf{S}$. The two
frames pictured here are not inertial because the direction of $\mathbf{L}$ changes together with the spins to conserve
$\mathbf{J}$.  These angles are defined in Eqs.~(\ref{t1t2dphi}), (\ref{JLeq}) and (\ref{varphidefinition}).}
\label{twoframes}
\end{figure}

The three angular momenta $\mathbf{L}$, $\mathbf{S_1}$ and $\mathbf{S_2}$ in principle constitute a nine-dimensional
parameter space.  However, there exist numerous constraints on the evolution of these parameters, greatly reducing the
number of degrees of freedom.  At the PN order considered here, the magnitudes of both spins are conserved throughout
the inspiral (see e.g. Ref.~\cite{2006PhRvD..74j4005B}), reducing the number of degrees of freedom from nine to seven.  
The magnitude of the orbital angular 
momentum is conserved on the precession time (although it shrinks on the radiation-reaction time), further reducing the
number of degrees of freedom from seven to six.  The total angular momentum $\mathbf{J} = \mathbf{L} + \mathbf{S_1} +
\mathbf{S_2}$ is also conserved on the precession time, reducing the number of degrees of freedom from six to three.  As
described in greater detail in the next subsection, the projected effective spin $\xi$ \cite{2001PhRvD..64l4013D,
2008PhRvD..78d4021R} is also conserved by both the orbit-averaged spin-precession equations at 2PN and radiation
reaction at 2.5 PN order, providing a final constraint that reduces the system to just two degrees of freedom.  In an appropriately
chosen non-inertial reference frame precessing about $\mathbf{J}$, precessional motion associated with one of these
degrees of freedom can be suppressed, 
  implying that the relative orientations of the three angular momenta $\mathbf{L}$,
$\mathbf{S_1}$ and $\mathbf{S_2}$ can be specified by just a single coordinate!  We will provide an explicit analytic
construction of this procedure in this and the following subsection.

We begin by introducing two alternative reference frames in which the relative orientations of the three angular momenta
can be specified explicitly.  As shown in the left panel of Fig.~\ref{twoframes}, one may choose the $z'$-axis to lie along
$\mathbf{L}$, the $x'$-axis such that $\mathbf{S_1}$ lies in the $x'z'$-plane, and the $y'$-axis to complete the orthonormal
triad.  In this frame only three independent coordinates are needed to describe the relative orientations of the angular
momenta; we choose them to be the angles
\begin{subequations}
\label{t1t2dphi}
\begin{align}
\cos\theta_1 &=\mathbf{\hat S_1} \cdot \mathbf{\hat L}\,,
\\
 \cos\theta_2&=\mathbf{\hat S_2} \cdot \mathbf{\hat L}\,,
\\
\cos\Delta\Phi&=\frac{\mathbf{\hat S_1} \times \mathbf{\hat L}}{|\mathbf{\hat S_1} \times  \mathbf{\hat L} |} \cdot 
\frac{\mathbf{\hat S_2} \times \mathbf{\hat L}}{|\mathbf{\hat S_2} \times \mathbf{\hat L} |},
\end{align}
where the sign of $\Delta\Phi$ is given by  (cf. Fig.~\ref{twoframes})
\begin{align}
\sign\Delta\Phi = \sign\{ \mathbf{L} \cdot [(\mathbf{S_1} \times \mathbf{L}) \times (\mathbf{S_2} \times \mathbf{L})] \}.
\end{align}
\end{subequations}

The relative orientations of the three angular momenta can alternatively be specified in a frame aligned with the total
angular momentum $\mathbf{J}$.  For fixed values of $L$, $S_1$, and $S_2$,  the allowed range for $J=|\mathbf{J}|$ is
\begin{subequations} 
\begin{align}
 J_{\rm min} \leq J \leq J_{\rm max}
\end{align}
where 
\begin{align}
J_{\rm min} &= \max(0 , L - S_1 - S_2, |S_1 - S_2| - L)\;, \label{Jmineq} \\
J_{\rm max}&= L+S_1+S_2\;.
\end{align}
\label{Jlim}\end{subequations}
As shown in the right panel of Fig.~\ref{twoframes}, one can choose the $z$-axis parallel to $\mathbf{J}$ and the $x$-axis
such that $\mathbf{L}$ lies in the $xz$-plane:
\begin{align}
\mathbf{J}= J  \hat{\mathbf{z}}\; \quad {\rm and} \quad  \mathbf{L} = L  \sin\theta_L\hat{\mathbf{x}} + L \cos \theta_L
\hat{\mathbf{z}} \;.
\label{JLeq}
\end{align}
The third unit vector $\hat{\mathbf{y}} = \hat{\mathbf{z}} \times \hat{\mathbf{x}}$ completes the orthonormal triad.   The total
spin $\mathbf{S} = \mathbf{S}_1 + \mathbf{S}_2 = \mathbf{J}-\mathbf{L}$ will also lie in the $xz$-plane:
\begin{equation}\label{bigSvector}
\mathbf{S} = -L\sin\theta_L\hat{\mathbf{x}} + (J-L\cos\theta_L)\hat{\mathbf{z}}~,
\end{equation}
implying
\begin{align}
\cos\theta_L &= \frac{J^2 + L^2 - S^2}{2JL}\;. \label{E:cstL}
\end{align}
We can also define a unit vector 
\begin{align}
\hat{\mathbf{S}}_\perp = \frac{ (J - L \cos\theta_L)\hat{\mathbf{x}}  + L \sin\theta_L \hat{\mathbf{z}} }{S} \label{E:Sperp}
\end{align}
which also lies in the $xz$-plane but is orthogonal to $\hat{\mathbf{S}}$.

While the magnitudes $L$ and $J$ of the orbital and total angular momenta are conserved on the precession timescale,
the same is not true for the total-spin magnitude $S$, which oscillates within the range
\begin{subequations} 
\begin{equation}
 S_{\rm min}  \leq S \leq S_{\rm max}  \;, 
\end{equation} 
where 
\begin{align}  \label{E:Smin}
S_{\rm min} &=  \max(|J-L|, |S_1 - S_2|) \;, \\
S_{\rm max} &= \min(J+L, S_1 + S_2) \;. \label{E:Smax}
\end{align}
\label{Slim}
\end{subequations}
$S$ can be used as a generalized coordinate to specify the directions of the angular momenta $\mathbf{J}$, $\mathbf{L}$,
and $\mathbf{S}$; we can see from Eqs.~(\ref{JLeq}) - (\ref{E:cstL}) that it is the only coordinate needed to specify these
directions in the $xyz$-frame.

Specifying the directions of the individual spins $\mathbf{S_1}$ and $\mathbf{S_2}$ in the $xyz$-frame requires an
additional generalized coordinate, which can be chosen to be the angle $\varphi'$ between $\hat{\mathbf{S}}_\perp$ in
Eq.~(\ref{E:Sperp}) and the projection of $\mathbf{S_1}$ into the plane spanned by $\hat{\mathbf{S}}_\perp$ and
$\hat{\mathbf{y}}$, as shown in the right panel of Fig.~\ref{twoframes}.  This angle corresponds to rotations of
$\mathbf{S_1}$ and $\mathbf{S_2}$ about $\mathbf{S}$ and is given analytically by
\begin{align}
\label{varphidefinition}
\cos\varphi' =  \frac{ \mathbf{\hat S_1} \cdot \mathbf{\hat S_\perp}}{  |\mathbf{\hat S_1} \times \mathbf{\hat S}|}~.
\end{align}
In terms of the two coordinates $S$ and $\varphi'$ varying on the precession timescale, the three angular momenta in the
$xyz$-frame are
\begin{subequations} 
\label{momenta}
\begin{align}
\mathbf{L} &=  \frac{A_1 A_2 }{2J}  \hat{\mathbf{x}} +  \frac{J^2
 + L^2 - S^2}{2J} \hat{\mathbf{z}}\,, \\
\mathbf{S_1} &=
 \frac{S^2 + S_1^2 - S_2^2}{2S} \hat{\mathbf{S}} + \frac{A_3 A_4}{2S}(\cos\varphi'\hat{\mathbf{S}}_\perp + \sin\varphi'
 \hat{\mathbf{y}})\notag \\
&= \frac{1}{4JS^2} [ -(S^2 + S_1^2 - S_2^2) A_1 A_2 
\notag \\ &\qquad + (J^2 - L^2 +S^2) A_3 A_4 \cos\varphi' ]
\hat{\mathbf{x}} \notag \\
& \quad + \frac{1}{2S}A_3 A_4 \sin\varphi' \hat{\mathbf{y}} \notag \\
& \quad + \frac{1}{4JS^2} [ (S^2 + S_1^2 - S_2^2)(J^2 - L^2 +S^2
)  
\notag \\ &\qquad + A_1 A_2 A_3 A_4\cos\varphi' ] \hat{\mathbf{z}}\,,\label{E:S1JLfr} 
\\
\mathbf{S_2} &=
\frac{S^2 + S_2^2 - S_1^2}{2S} \hat{\mathbf{S}} - \frac{A_3 A_4}{2S}(\cos\varphi'\hat{\mathbf{S}}_\perp + \sin\varphi'
\hat{\mathbf{y}})
\notag \\
&= -\frac{1}{4JS^2} [ (S^2 + S_2^2 - S_1^2)A_1 A_2 \notag \\
&\qquad+ (J^2 - L^2 +S^2) A_3 A_4\cos\varphi' ]
\hat{\mathbf{x}} \notag \\
& \quad - \frac{1}{2S}A_3 A_4 \sin\varphi'\hat{\mathbf{y}} \notag \\
& \quad + \frac{1}{4JS^2} [ (S^2 + S_2^2 - S_1^2)(J^2 - L^2 +S^2
) \notag \\
&\qquad  - A_1 A_2 A_3 A_4\cos\varphi' ] \hat{\mathbf{z}} \label
{E:S2JLfr}\,,
\end{align}
\end{subequations}
where we defined:
\begin{subequations}
\begin{align}
A_1 &\equiv [J^2 - (L - S)^2]^{1/2}\,, \\
A_2 &\equiv [(L + S)^2 - J^2]^{1/2}\,, \\
A_3 &\equiv [S^2 - (S_1 - S_2)^2]^{1/2}\,, \\
A_4 &\equiv [(S_1 + S_2)^2 - S^2]^{1/2}\,.
\end{align}
\label{Ais}\end{subequations} 
All the $A_i$'s are real and positive in the ranges specified by Eqs.~(\ref{Jlim}) and (\ref{Slim}).

\subsection{Effective potentials and resonances}
\label{subseceffpot}

\begin{figure*}[thb] 
\includegraphics[width=\textwidth]{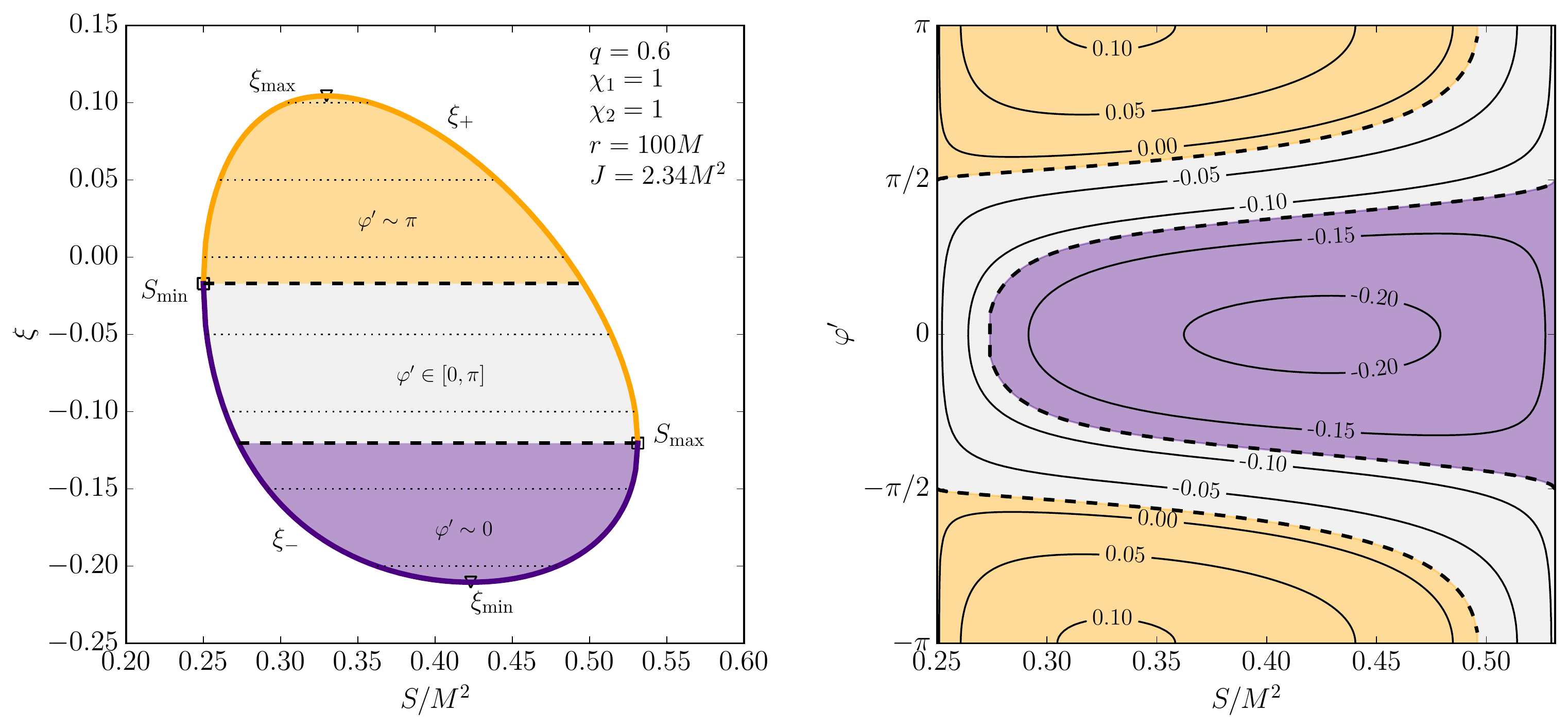} 
\caption{{\it Left:} Effective potentials $\xi_\pm(S)$ for BBHs with $q=0.6$, $\chi_1=\chi_2=1$, $r=100M$, and
$J=2.34 M^2$.  Conservation of the projected effective spin $\xi$ constrains the BBH spins to precess along horizontal
lines bounded by the effective-potential curves.  The horizontal dashed lines intersecting the effective potentials at
$S_{\rm min}$ and $S_{\rm max}$ (marked by empty squares) divide BBH spin precession into three different
morphological phases distinguished by whether the angle $\varphi'$ defined by Eq.~(\ref{varphidefinition}) oscillates about
$\pi$ (top orange region), circulates from 0 to $2\pi$ (middle grey region), or oscillates about 0 (bottom purple region).  The effective
potentials admit two extrema $\xi_{\rm min}$ and $\xi_{\rm max}$ (marked by empty triangles) corresponding to the
spin-orbit resonances discovered in Ref.~\cite{2004PhRvD..70l4020S}. {\it Right:}  Contours of constant $\xi(S, \varphi')$ given
by Eq.~(\ref{xiSvaphi}) for the same binary parameters.  As BBH spins precess along the horizontal dashed lines in the left panel, they move along the
curves in the $S\varphi'$-plane in the right panel illustrating the three morphological phases.}  \label{varphi_evolution}
\end{figure*}

As anticipated in the previous subsection, there is an additional conserved quantity that can be used to eliminate $\varphi'$
and thereby specify $\mathbf{L}$, $\mathbf{S_1}$, and $\mathbf{S_2}$ with the single generalized coordinate $S$.  This
quantity is the projected effective spin \cite{2001PhRvD..64l4013D,2008PhRvD..78d4021R}
\begin{equation} \label{E:xidef}
\xi \equiv M^{-2} [(1+q)\mathbf{S}_1 + (1+q^{-1})\mathbf{S}_2] \cdot \hat{\mathbf{L}}
\end{equation}
which is a constant of motion of the orbit-averaged spin-precession equations at 2PN order and is also conserved by
radiation reaction at 2.5 PN order. Using Eqs.~(\ref{momenta}), we can express $\xi$ as a function of $S$ and $\varphi'$
\begin{align} \label{xiSvaphi}
\xi(S, \varphi^\prime) &= \{ (J^2 - L^2 - S^2)[S^2(1+q)^2 - (S_1^2 - S_2^2)(1- q^2)] \notag \\
& \quad - (1- q^2) A_1A_2A_3A_4\cos\varphi^\prime \}/(4qM^2S^2L)\,.
\end{align}
Conservation of $\xi$ restricts binary evolution to one-dimensional curves $\xi(S, \varphi^\prime)=\xi$ in the $S\varphi'$-plane as shown in the right panel of Fig.~\ref{varphi_evolution}.  The simple dependence of $\xi(S, \varphi^\prime)$ on
$\varphi^\prime$ motivates us to define two ``effective potentials'' \cite{2015PhRvL.114h1103K} corresponding to the
extreme cases $\cos\varphi'= \mp 1$ for which $\mathbf{L}$, $\mathbf{S_1}$ and $\mathbf{S_2}$ are all coplanar:
\begin{align} \label{effpot}
\xi_{\pm}(S) &= \{ (J^2 - L^2 - S^2)[S^2(1+q)^2 - (S_1^2 - S_2^2)(1- q^2)] \notag \\
& \quad \pm (1- q^2) A_1A_2A_3A_4 \}/(4qM^2S^2L)\,.
\end{align}
At $S_{\rm min}$ and $S_{\rm max}$ 
\begin{align}
\xi_{-}(S_{\rm min}) = \xi_{+}(S_{\rm min})\;, \qquad \xi_{-}(S_{\rm max}) = \xi_{+}(S_{\rm max})\;,
\end{align}
because one of the $A_i$'s defined in Eqs.~(\ref{Ais}) vanishes if $S=S_{\rm min}$ or $S=S_{\rm max}$.  The functions
$\xi_{\pm}(S)$ thus form a loop that encloses all allowed values of $S$ and $\xi$, as shown in the left panel
of Fig.~\ref{varphi_evolution}.  BBHs are constrained to evolve back and forth along horizontal line segments of constant
$\xi$ bounded by the two effective potentials $\xi_\pm(S)$.
The turning points in the evolution of $S$ are given by the solutions of $\xi_{\pm}(S) = \xi$, where the binary meets an
effective potential. Once squared, the equation $\xi_{\pm}(S) = \xi$ reduces to the following cubic equation in $S^2$:
\begin{subequations}
\begin{align}
\sigma_6 S^6 + \sigma_4 S^4 + \sigma_2 S^2 +\sigma_0=0 \,,
\end{align}
where 
\begin{align}
\sigma_6 &=  q (1+q)^2 \,,\\
\sigma_4 &=  (1+q)^2 [-2J^2 q+L^2 \left(1+ q^2\right)+2 L M^2 \xi  q \notag\\ &\quad 
+ (1-q) \left(S_2^2- q S_1^2\right)]\,, \\
\sigma_2 &= 
2 (1 + q)^2 
  (1 - q) [ J^2 (q S_1^2 - S_2^2) 
\notag\\ &\quad 
  - L^2 (S_1^2 - q S_2^2) ]  
 +  q  (1 + q)^2 (J^2 - L^2)^2
\notag\\ &\quad 
 - 2 L M^2  \xi  q (1 + 
   q) [(1 + q)  (J^2 - L^2) 
\notag\\ &\quad 
   + (1 - q) (S_1^2 - S_2^2)]  
+4 L^2  M^4 \xi^2  q^2 \,,
\\
\sigma_0 &=      (1 - q^2)
   [ L^2 (1 - q^2) (S_1^2 - S_2^2)^2 
\notag\\ &\quad 
 - (1 + q) (q S_1^2 - S_2^2) (J^2 - L^2)^2
\notag\\ &\quad 
 + 2 L   M^2  q  \xi (S_1^2 - S_2^2)  (J^2 - L^2)
  ]\,,
\end{align}
\end{subequations}
which admits at most three real solutions for $S>0$.  The number of solutions in the range allowed by Eqs.~(\ref{Slim}) must
be even because the two effective potentials form a closed loop and the Jordan curve theorem requires the number of
intersections between a continuous closed loop and a line to be even \cite{hatcher2002algebraic} (although these
intersections can coincide at extrema).
Since two is the largest even number less than three, the equation $\xi_{\pm}(S) = \xi$ will generally have two solutions 
which we denote by $S_{\pm}$ ($S_-\leq S_+$).

The total-spin magnitude $S$ will oscillate between $S_-$ and $S_+$  implying that spin precession is regular or quasi-periodic 
(this will be shown explicitly in Sec.~\ref{timedependent} below). 
The motion of the spins is not fully periodic because in an inertial frame the basis vectors $\hat{\mathbf{x}}$
and $\hat{\mathbf{y}}$ will generally not precess about $\mathbf{J}$ by a rational multiple of $\pi$ radians in the time it
takes $S$ to complete a full cycle from $S_-$ and $S_+$ and back again.  The turning points $S=S_\pm$ lie on the
effective potentials, implying from the definition $\cos\varphi' = \mp1$ that all three vectors $\mathbf{L}$, $\mathbf{S_1}$,
and $\mathbf{S_2}$ are coplanar.  The qualitative evolution of $\varphi'$ is related to the nature of the turning points $S_{\pm}$. This is illustrated in Fig.~\ref{varphi_evolution}, where horizontal lines in the effective-potential diagram (left panel) correspond to contours of constant $\xi(S,\varphi')$, computed using Eq.~(\ref{xiSvaphi}) (right panel).  Three different cases are possible.

\begin{subequations}
\begin{enumerate}
\item Both turning points lie on $\xi_+$:
\begin{align}
\xi_+(S_+)=\xi_+(S_-)=\xi\,.
\end{align}
$\varphi'$ oscillates about $\pi$ never reaching $0$ (orange region in Fig.~\ref{varphi_evolution}).

\item One turning point is on  $\xi_-$ and the other  is on $\xi_+$:
\begin{align}
\xi_\pm(S_-)=\xi_\mp(S_+)=\xi\,.
\end{align}
$\varphi'$ monotonically circulates from $-\pi$ to $\pi$ during each precession cycle (grey region in
Fig.~\ref{varphi_evolution}).

\item Both turning points lie on $\xi_-$:
\begin{align}
\xi_-(S_+)=\xi_-(S_-)=\xi\,.
\end{align}
$\varphi'$ oscillates about $0$ never reaching $\pi$ (purple region in Fig.~\ref{varphi_evolution}).
\end{enumerate}
\end{subequations}
The boundaries between the three regions are given by those values of $\xi$ at which one of the turning points $S_\pm$
coincides with either $S_{\rm min}$ or $S_{\rm max}$ (dashed lines in Fig.~\ref{varphi_evolution}).
Note that $\xi(S_{\rm min})$ may be less or greater than  $\xi(S_{\rm max})$ depending on the values of $q$, $\chi_i$, $r$ and $J$.  

The two turning points are degenerate $(S_{+}=S_{-})$ at the extrema $\xi_{\rm min}$ and $\xi_{\rm max}$ of the effective
potentials.  At these extrema the derivatives
\begin{align}\label{dereffpot}
\frac{d\xi_\pm}{dS} &= \frac{1+q}{2qM^2S^3L} \bigg\{ (1-q)(J^2-L^2)(S_1^2 - S_2^2) - (1+q)S^4 
\notag \\ 
&\pm  \frac{1-q}{A_1A_2A_3A_4}  \Big[ S^8 - (J^2 + L^2 + S_1^2 + S_2^2) S^6  
\notag \\ 
&+ (J^2 + L^2)(S_1^2 - S_2^2)^2 S^2  
+ (S_1^2 + S_2^2)(J^2 - L^2)^2 S^2  
\notag \\ 
&-  (S_1^2 - S_2^2)^2(J^2 - L^2)^2 \Big] \bigg\}
\end{align}
vanish and $S=S_-=S_+$ is constant. Since
\begin{subequations}
\begin{align}
\lim_{S \to S_{\rm min}} \frac{d\xi_+}{dS} \geq \lim_{S \to S_{\rm min}} \frac{d\xi_-}{dS}\,,
\\
 \lim_{S \to S_{\rm max}} \frac{d\xi_+}{dS} \leq \lim_{S \to S_{\rm max}} \frac{d\xi_-}{dS}\,,
 \end{align}
\end{subequations}
and at most two turning points can exist, it follows that $\xi_+$ admits a single maximum  in $[S_{\rm min}, S_{\rm max}]$
and $\xi_-$ admits a single minimum in $[S_{\rm min}, S_{\rm max}]$.  The effective potentials therefore have exactly two
distinct extrema for each value of the constants $J$, $r$, $q$, $\chi_1$ and $\chi_2$.  As clarified below, these special
configurations correspond to the spin-orbit resonances discovered by other means in Ref.~\cite{2004PhRvD..70l4020S}.

The equal-mass limit $q\to 1$ corresponds to $\xi_+(S)=\xi_-(S)$ [cf. Eq.~(\ref{effpot})]  implying that $S$ is constant for all
values of $\xi$ [note that $\xi_\pm(S_{\rm min})\neq \xi_\pm(S_{\rm max})$]. This fact was noted at least as early as 2008 by
Racine~\cite{2008PhRvD..78d4021R} and it was recently exploited in numerical-relativity simulations
\cite{2014PhRvD..89j4052L,2015PhRvL.114n1101L}, but the constancy of $S$ is a peculiarity of the equal-mass case and
does not hold for generic binaries.

\subsection{Morphological classification}
\label{subsecmorph}
 
Although the evolution of $\varphi'$ already provides a way to characterize the precessional dynamics
(Fig.~\ref{varphi_evolution}), a more intuitive understanding can be gained by switching back to the $\mathbf{L}$-aligned
frame illustrated in the left panel of Fig.~\ref{twoframes}.  Substituting Eqs.~(\ref{momenta}) and (\ref{xiSvaphi}) into Eq.~(\ref{t1t2dphi}), we can express the angles $\theta_1$, $\theta_2$ and $\Delta\Phi$ as functions of $S$, $J$ and $\xi$. This
yields the remarkably simple expressions \cite{2015PhRvL.114h1103K}
\begin{subequations} \label{t1t2dphisolutions}
\begin{align}
\cos\theta_1 &=  \frac{1}{2(1-q)S_1} \left[ \frac{J^2 - L^2 -S^2}{L} - \frac{2qM^2\xi}{1+q} \right]\,, \label{E:cs1} \\
\cos\theta_2 &=  \frac{q}{2(1-q)S_2} \left[ -\frac{J^2 - L^2 -S^2}{L} + \frac{2M^2\xi}{1+q} \right]\,, \label{E:cs2} \\
\cos\Delta\Phi &= \frac{\cos\theta_{12} - \cos\theta_1\cos\theta_2}{\sin\theta_1\sin\theta_2} \,,
\label{E:dphi}
\end{align}
where the angle $\theta_{12}=\arccos \mathbf{\hat S_1}\cdot \mathbf{\hat S_2}$ between the two spins can also be written in terms of $S$:
\begin{equation} 
\cos\theta_{12} = \frac{S^2 - S_1^2 - S_2^2}{2S_1S_2}.
\end{equation}
\end{subequations}

\begin{figure}
\begin{center}
\includegraphics[width=0.85\columnwidth]{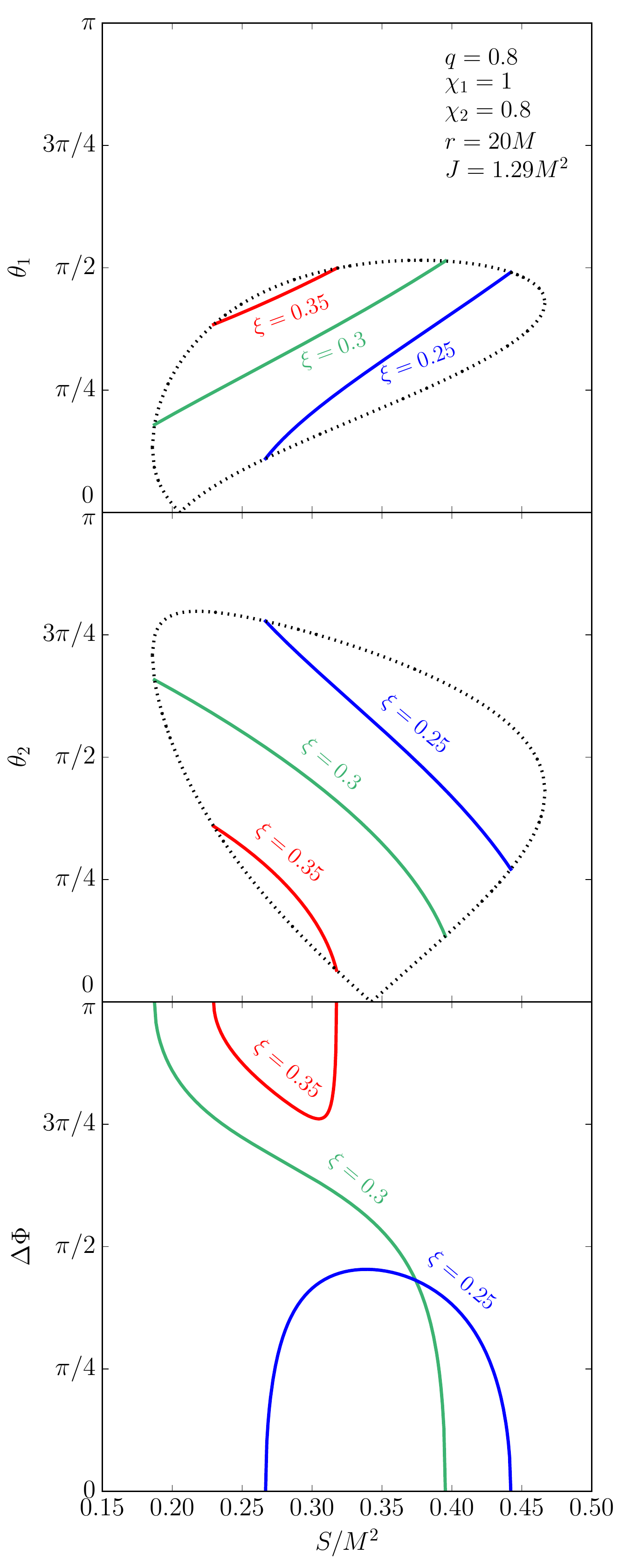}
\caption{Analytical solutions given by Eq.~(\ref{t1t2dphisolutions}) for the evolution of the angles $\theta_1$ (top panel),
$\theta_2$ (middle panel), and $\Delta\Phi$ (bottom panel) during a precession cycle.  The evolution of three binaries with
$\xi=0.25$ (blue), $0.3$ (green) and  $0.35$ (red) is shown for $q=0.8$, $\chi_1=1$, $\chi_2=0.8$, $r=20M$ and
$J=1.29 M^2$.  The evolution of $\theta_1$ and $\theta_2$ is monotonic during each half of a precession cycle and is
bounded by the dotted lines for which $\cos\varphi = \mp1$ [these curves can be found by substituting $\xi_\pm(S)$ for
$\xi$ in Eq.~(\ref{t1t2dphisolutions})].  Three classes of solutions are possible and define the binary morphology: 
$\Delta\Phi$ can oscillate about 0 ($\xi =0.25$), circulate ($\xi=0.3$) or oscillate about $\pi$ ($\xi=0.35$).  An animated
version of this figure is available online at Ref.~\cite{DGwebsite}, where precession solutions are evolved on $t_{\rm RR}$.}
\label{angle_solutions}
\end{center}
\end{figure}

\begin{figure*}[p]
\centering
\includegraphics[width=0.95\textwidth]{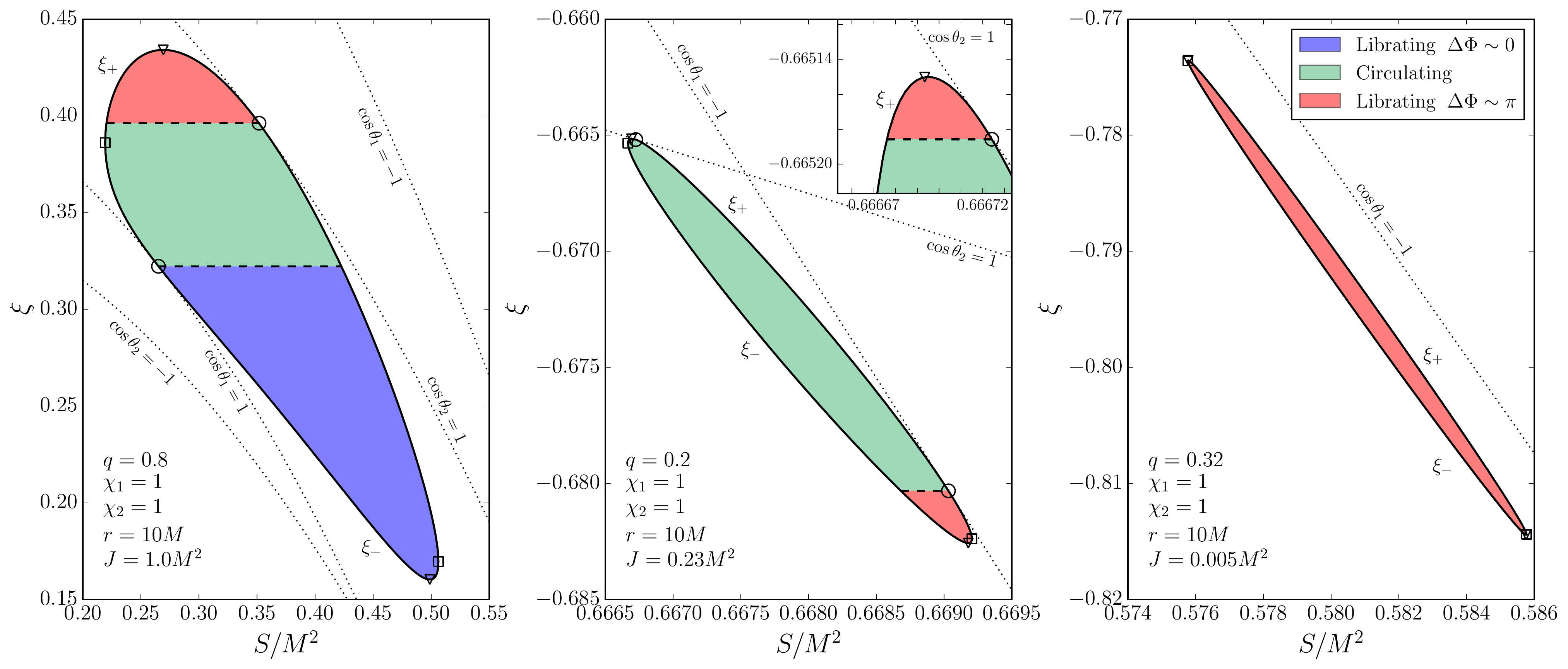}
\caption{Effective potentials $\xi_\pm(S)$ of Eq.~(\ref{effpot}) for values of $L$, $J$, $S_1$, and $S_2$ leading to three
different sets of spin morphologies. The loop formed by the two curves encloses all allowed configurations for the constants
listed in the legends.  As in the left panel of Fig.~\ref{varphi_evolution}, empty squares mark the extrema of
$S$ ($S_{\rm min}$ and $S_{\rm max}$), empty triangles mark the extrema of $\xi$ ($\xi_{\rm min}$ and $\xi_{\rm max}$),
and conservation of $\xi$ restricts the BBH spins to precess along horizontal lines between the turning points $S_\pm$.
BBH spin precession can be classified into three different morphologies by the behavior of $\Delta\Phi$ during a
precession cycle: oscillation about 0 (blue region), circulation from $-\pi$ to $\pi$ (green region), or oscillation about $\pi$
(red region).  The dashed boundaries between these morphologies occur at values of $\xi$ where the dotted curves
$\cos\theta_i=\pm 1$ intersect the effective-potential loop, as shown by the empty circles.  All three morphologies are
present if one intersection occurs on $\xi_+(S)$ and a second occurs on $\xi_-(S)$ (left panel), oscillation of $\Delta\Phi$
about 0 is forbidden if two intersections occur on either $\xi_+(S)$ or $\xi_-(S)$ (middle panel), and only oscillations about
$\pi$ are allowed if there are no such intersections (right panel).}
\label{threeeffpot}
\vspace{1cm}
\includegraphics[width=0.95\textwidth]{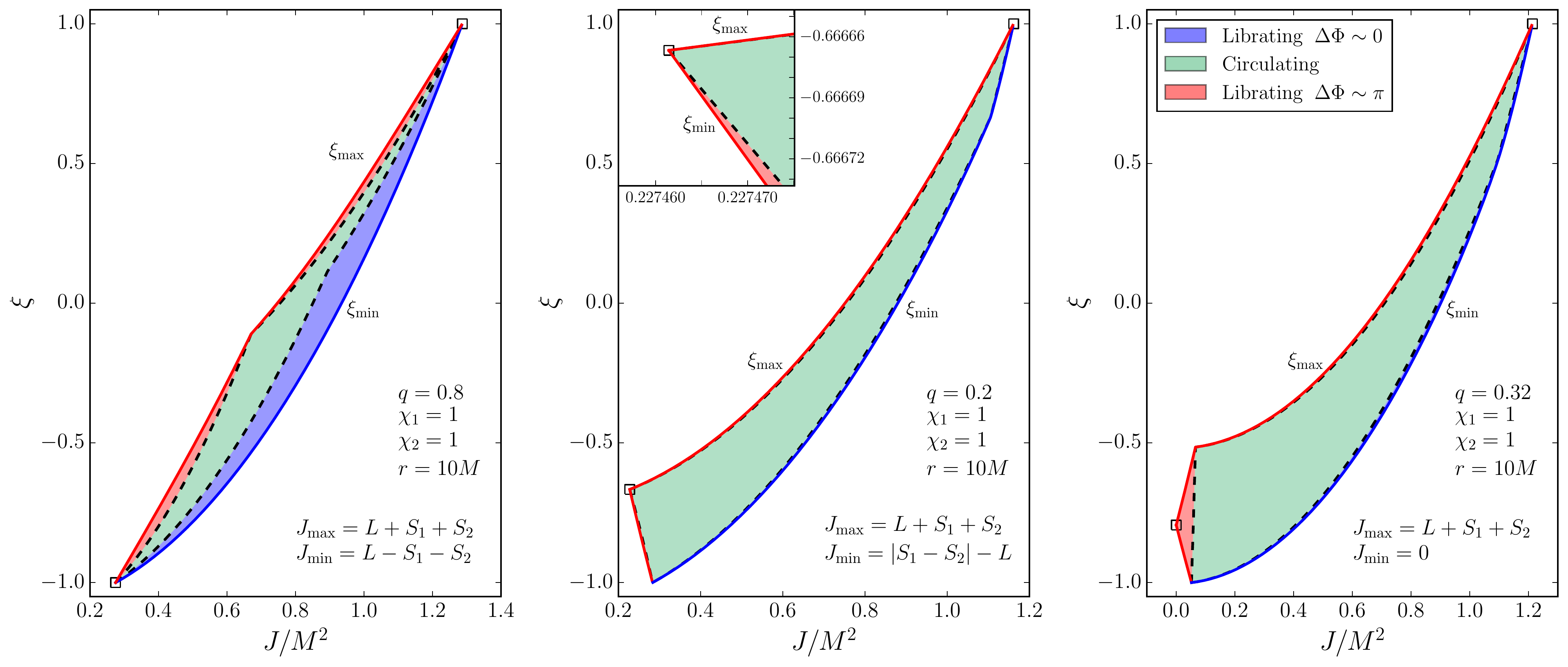}
\caption{The ($J,\xi$) parameter space for BBHs with different minimum allowed total angular momentum $J_{\rm min}$. 
BBH spin morphology is shown with different colors, as indicated in the legend.  The extrema $\xi_{\rm min}(J)$ and
$\xi_{\rm max}(J)$ of the effective potentials constitute the edges of the allowed regions and are marked by solid blue (red)
curves for $\Delta\Phi=0~(\pi)$.  Dashed lines mark the boundaries between the different morphologies. The parameters
$q$, $\chi_1$, $\chi_2$ and $r$ are chosen as in Fig.~\ref{threeeffpot}, whose panels can be thought of as vertical
(constant $J$) ``sections'' of this figure (where we suppress the $S$ dependence). The lowest allowed value of $\xi$
occurs at $J=|L-S_1-S_2|$ in all three panels.  Three phases are present for each vertical section with $J>|L-S_1-S_2|$.
This condition may either cover the entire parameter space (left panel) or leave room for additional regions where vertical
sections include two different phases in which $\Delta\Phi$ oscillates about $\pi$ and a circulating phase in between 
(center panel) or only a single phase where the spins librate about $\Delta\Phi = \pi$ (right panel).  An animated version of this figure evolving on the radiation-reaction time $t_{\rm RR}$ is available online~\cite{DGwebsite}.}
\label{threeJxi}
\end{figure*}

Equations~(\ref{t1t2dphisolutions}) parametrize double-spin binary precession using a single parameter $S$.
Some examples of the evolution of these angles over a precessional cycle are given in Fig.~\ref{angle_solutions}. The
evolution of  $\theta_1$ and $\theta_2$ is monotonic as $S$ evolves between its two turning points $S_\pm$; over a full
precessional cycle these angles oscillate between two extrema lying on the effective potentials (dotted curves in
Fig.~\ref{angle_solutions}).  The evolution of $\Delta\Phi$ can be classified into three morphological phases similar to that
of $\varphi'$:
\begin{subequations}
\begin{enumerate} \label{E:MorphCrit}
\item   $\Delta\Phi$ oscillates about $0$ (never reaching $\pi$) if 
\begin{align}
\Delta\Phi(S_-) = \Delta\Phi(S_+) = 0 \,;
\end{align}
 \item  $\Delta\Phi$ circulates through the  full range $[-\pi,\pi]$ if
 \begin{align}
\Delta\Phi(S_\pm) = 0 \quad {\rm and}\quad \Delta\Phi(S_\mp) = \pi\,;
\label{dphicirculation}
\end{align}
\item $\Delta\Phi$ oscillates about $\pi$ (never reaching $0$) if
 \begin{align}
\Delta\Phi(S_-) = \Delta\Phi(S_+) = \pi\,.
\end{align}
\end{enumerate}
 \end{subequations}
The evolution of $\Delta\Phi$ allows us to unambiguously categorize the precessional dynamics into the three different
classes listed above.  We refer to these classes as \emph{morphologies} because of the different shapes traced out by the
BBH spins over a precession cycle.  We show some examples of how the allowed region inside the effective-potential loop
is divided between these three morphologies in Fig.~\ref{threeeffpot}.
BBHs in the two oscillating morphologies are adjacent to the extrema of the effective potentials ($\xi_{\rm min}$ and
$\xi_{\rm max}$), while circulating binaries (if present) fill the gap in between.  Schnittman's spin-orbit resonances
\cite{2004PhRvD..70l4020S} can be reinterpreted as the limits of the two oscillating morphologies when the
``precessional amplitude'' $S_+-S_-$ goes to zero at $\xi_{\rm min}$ and $\xi_{\rm max}$, much like how circular orbits are
the limits of eccentric orbits as the amplitude of the radial oscillations goes to zero.

According to the criteria listed in Eqs.~(\ref{E:MorphCrit}), boundaries between the three morphologies (shown by horizontal
dashed lines in Fig.~\ref{threeeffpot}) occur at values of $\xi$ where $\cos\Delta\Phi$ given by Eq.~(\ref{E:dphi}) changes
discontinuously at one of the turning points $S_\pm$ along the effective-potential loop $\xi_\pm(S)$.  We know that
$\Delta\Phi$ is either $0$ or $\pi$ along $\xi_\pm(S)$ because $\mathbf{L}$, $\mathbf{S}_1$, and $\mathbf{S}_2$ are
coplanar when $\cos\varphi' = \pm1$ (see Fig.~\ref{twoframes}).  A discontinuity can only occur when the denominator of
Eq.~(\ref{E:dphi}) vanishes, i.e. where one of the spins is either aligned or anti-aligned with the orbital angular momentum
($\sin\theta_i = 0$).  These discontinuities can only happen at the turning points $S_\pm$ because of the monotonic
evolution of $\theta_i$ during each half of the precession cycle, as shown in the top and middle panels of
Fig.~\ref{angle_solutions}.  The four contours $\cos\theta_i = \pm1$ ($\sin\theta_i = 0$) are shown by dotted curves in
Fig.~\ref{threeeffpot}; we see that a boundary between morphologies occurs whenever these curves are tangent to the
effective-potential loop $\xi_\pm(S)$.  These boundaries had previously been described as unstable resonances
\cite{2004PhRvD..70l4020S}.

The geometrical constraints imposed by Eqs.~(\ref{Jlim}) and (\ref{Slim}) imply that some morphologies may not be allowed
for given values of $L,J,q,\chi_1$, and $\chi_2$.  Three qualitatively different scenarios can occur, exemplified by the three
panels of Fig.~\ref{threeeffpot}:
\begin{enumerate}

\item {\it Left panel:} BBH spins precess in all three of the morphologies listed in Eq.~(\ref{E:MorphCrit}).  Libration about the
coplanar configuration $\Delta\Phi = 0$ occurs for values of $\xi$ close to $\xi_{\rm min}$, libration about the
$\Delta\Phi = \pi$ configuration is found near $\xi_{\rm max}$, and $\Delta\Phi$ circulates for intermediate values of
$\xi$.  Our analysis in Ref.~\cite{2015PhRvL.114h1103K} was restricted to this case.

\item {\it Middle panel:} $\Delta\Phi$ oscillates about $\pi$ for $\xi$ close to both $\xi_{\rm min}$ and $\xi_{\rm max}$, with
circulation still allowed for intermediate values of $\xi$.

\item {\it Right panel:} $\Delta\Phi$ oscillates about $\pi$ for all values $\xi_{\rm min}< \xi < \xi_{\rm max}$ (circulation and
oscillation about 0 are both forbidden).

\end{enumerate}

To distinguish these scenarios, it is useful to examine the values of $\Delta\Phi$ on the effective-potential loop at the
extrema $\xi_{\rm min}$ and $\xi_{\rm max}$.  Although it is straightforward to evaluate $\Delta\Phi$ numerically at
$\xi_{\rm max}$, one can gain more intuition by instead evaluating it at $S_{\rm min}$.  The value of $\Delta\Phi$ is the
same at these two points since the slope of the effective-potential loop $\xi_+(S)$ connecting them is positive while that of
the $\cos\theta_i = \pm1$ contours is negative (as can be seen in Fig.~\ref{threeeffpot}).  The curves therefore cannot be
tangent to each other implying that $\Delta\Phi$ must remain constant on this portion of the loop.  Equation~(\ref{E:Smin})
requires that $S_{\rm min}$ equals the greater of $|J-L|$ and $|S_1-S_2|$; in the former case $\mathbf{L}$ and
$\mathbf{J}$ are anti-aligned, while in the latter case $\mathbf{S}_1$ and $\mathbf{S}_2$ are anti-aligned.  In either case, the
components of $\mathbf{S}_1$ and $\mathbf{S}_2$ perpendicular to $\mathbf{L}$ are anti-aligned ($\Delta\Phi = \pi$). 
This implies that $\Delta\Phi$ will oscillate about $\pi$ near $\xi_{\rm max}$ for all values of $J$, $L$, $S_1$, and $S_2$
(as can be seen in all three panels of Fig.~\ref{threeeffpot}).

The values of $\Delta\Phi$ on the effective-potential loop at $\xi_{\rm min}$ and $S_{\rm max}$ are also the same because
the segment of the curve connecting them has a positive slope.  Equation (\ref{E:Smax}) indicates that $S_{\rm max}$ equals the
lesser of $|J+L|$ and $|S_1+S_2|$; in the former case $\mathbf{L}$ and $\mathbf{J}$ are anti-aligned, while in the latter
case $\mathbf{S}_1$ and $\mathbf{S}_2$ are aligned.  The former case again requires the components of $\mathbf{S}_1$
and $\mathbf{S}_2$ perpendicular to $\mathbf{L}$ to be anti-aligned ($\Delta\Phi = \pi$) but now the latter case requires
these components to be aligned ($\Delta\Phi = 0$).  For values of $J$, $L$, $S_1$, and $S_2$ for which this latter case
applies, $\Delta\Phi$ will oscillate about 0 near $\xi_{\rm min}$ and we have determined that all three morphologies are
possible, as shown in the left panel of Fig.~\ref{threeeffpot}.

To distinguish the remaining two scenarios (whether or not $\Delta\Phi$ circulates for intermediate values of $\xi$), we must
examine the intersections of the $\cos\theta_i = \pm1$ contours with the effective-potential loop $\xi_\pm(S)$.  There can be
either zero or two of such intersections.  If no intersections occur, $\Delta\Phi$ remains equal to $\pi$ around the entire loop 
just as it is at $\xi_{\rm max}$ and only oscillations about this value are possible, as shown in the right panel of
Fig.~\ref{threeeffpot}.  If there are two intersections, they must happen on the two portions of the loop with negative slopes
(the segment connecting $S_{\rm min}$ and $\xi_{\rm min}$ and the segment connecting $S_{\rm max}$ and
$\xi_{\rm max}$).  If both intersections happen on the {\it same} segment, $\Delta\Phi$ switches from $\pi$ to 0 and back
again as one traverses the loop from $\xi_{\rm max}$ to $\xi_{\rm min}$ resulting in the introduction of a circulating phase
before restoring oscillations about $\pi$ near $\xi_{\rm min}$, as seen in the middle panel of Fig.~\ref{threeeffpot}.  If the two
intersections happen on {\it different} segments, $\Delta\Phi$ switches to 0 at the first turning point and then to $\pi$ at the
other leading to oscillations about 0 near $\xi_{\rm min}$, as seen previously in the left panel of Fig.~\ref{threeeffpot}. 

To summarize, the number of allowed morphologies in the effective-potential diagrams of Fig.~\ref{threeeffpot} depends on
the magnitude of the total angular momentum $J$:
 \begin{enumerate}
 \item All three phases are allowed if 
 \begin{align}
 J > S_1 + S_2 - L\,.
 \end{align}
This condition implies $S_{\rm max} = S_1 + S_2$ and hence $\Delta\Phi(\xi_{\rm min})=0$ (Fig.~\ref{threeeffpot}, left panel).
\item For lower values of $J$ such that  
 \begin{align}
L - |S_1 - S_2| < J < S_1 + S_2 - L\,,
 \end{align}
$\Delta\Phi$ will oscillate about $\pi$ near $\xi_{\rm min}$ and $\xi_{\rm max}$ and circulate from $-\pi$ to $\pi$
for intermediate values of $\xi$ (Fig.~\ref{threeeffpot}, middle panel). The first inequality ensures that two (anti)aligned
configurations  ($\sin\theta_i=0$) can be found, while the second prevents $\Delta\Phi = 0$.
 \item Finally, for
 \begin{align}
 J < \min (S_1 + S_2 - L,L - |S_1 - S_2|)\,,
 \end{align}
the condition $\sin\theta_i=0$ cannot be satisfied and $\Delta\Phi$ must oscillate about $\pi$ (Fig.~\ref{threeeffpot},
right panel).
\end{enumerate}
Whether these conditions can be satisfied is determined by the limits on $J$ given by Eqs.~(\ref{Jlim}).  In particular,
$J_{\rm min} = L - S_1 - S_2$ is a sufficient but not necessary condition for all three morphologies to coexist, while
$J_{\rm min} = 0$ is a necessary but not sufficient condition for the single-phase case. The three-phase case
was considered in our {\it Letter} \cite{2015PhRvL.114h1103K} and is the only allowed case at sufficiently large binary
separations ($L > S_1 + S_2$).

The $J\xi$-plane shown in Fig.~\ref{threeJxi} shows all BBH spin configurations for fixed values of $q$, $\chi_1$, $\chi_2$
and $r$ at once.  Since $J$ and $\xi$ are constant on the precession time $t_{\rm pre}$, the position of BBHs in this figure
is fixed on this timescale.  The effective-potential diagrams of Fig.~\ref{threeeffpot} can be thought of as vertical sections of
Fig.~\ref{threeJxi} at fixed $J$ where the $S$ direction has been expanded.  Each panel of Fig.~\ref{threeJxi} refers to a
different choice of $J_{\rm min}$ from Eqs.~(\ref{Jmineq}).  $\Delta\Phi$ can only oscillate about 0 if
\mbox{$J> |L-S_1-S_2|$}.  From Eq.~(\ref{E:xidef}), the limit $J=|L-S_1-S_2|$ corresponds to the lowest allowed value of
$\xi$.  For separations large enough that $L>S_1+S_2$, this configuration also corresponds to $J_{\rm min}$ in which case
$\Delta\Phi$ can oscillate about 0 for all allowed values of $J$ (Fig.~\ref{threeJxi}, left panel).  If $L$ is sufficiently small to
admit values of $J$ such that $J<|L- S_1 - S_2|$, a new region of the parameter space where $\Delta\Phi=0$ is forbidden
appears at small $J$ (middle and right panels of Fig.~\ref{threeJxi}).  If even lower values $J < |S_1 - S_2| - L$ can be
reached (i.e., if $J_{\rm min}=0$), the leftmost region of the $J\xi$-plane does not even allow a circulating phase (right
panel of Fig.~\ref{threeJxi}).

The center and right panels of Fig.~\ref{threeJxi} reveal that the regions for which $\Delta\Phi$ oscillates (shown in blue
and red) are very small for $L<S_1+ S_2$.  This follows from the fact that these small values of the orbital angular
momentum can only be achieved in the PN regime ($r \gtrsim 10M$) for low mass ratios.  Oscillation of $\Delta\Phi$ relies
upon coupling between the two BBH spins, and the spin $\mathbf{S}_2$ becomes increasingly ineffective at maintaining
this coupling as $q \to 0$ (cf. Sec.~\ref{msasymmetry} below for more details).  Nonetheless, a small region of the
parameter space is always occupied by librating binaries as $\xi$ approaches the resonant values $\xi_{\rm min}$ and
$\xi_{\rm max}$.  For each value of $\xi$ (horizontal sections of Fig.~\ref{threeJxi}), one $\Delta\Phi=0$ resonance and one
$\Delta\Phi=\pi$ resonance occur at the largest ($\Delta\Phi=0$) and the lowest ($\Delta\Phi=\pi$) allowed values of $J$.
The effective spin $\xi$ is therefore a good parameter to identify the resonant solutions, as we pointed out in
Ref.~\cite{2014PhRvD..89l4025G}.

\subsection{Time dependence}
\label{timedependent}

\begin{figure}[t]
\begin{center}
\includegraphics[width=\columnwidth]{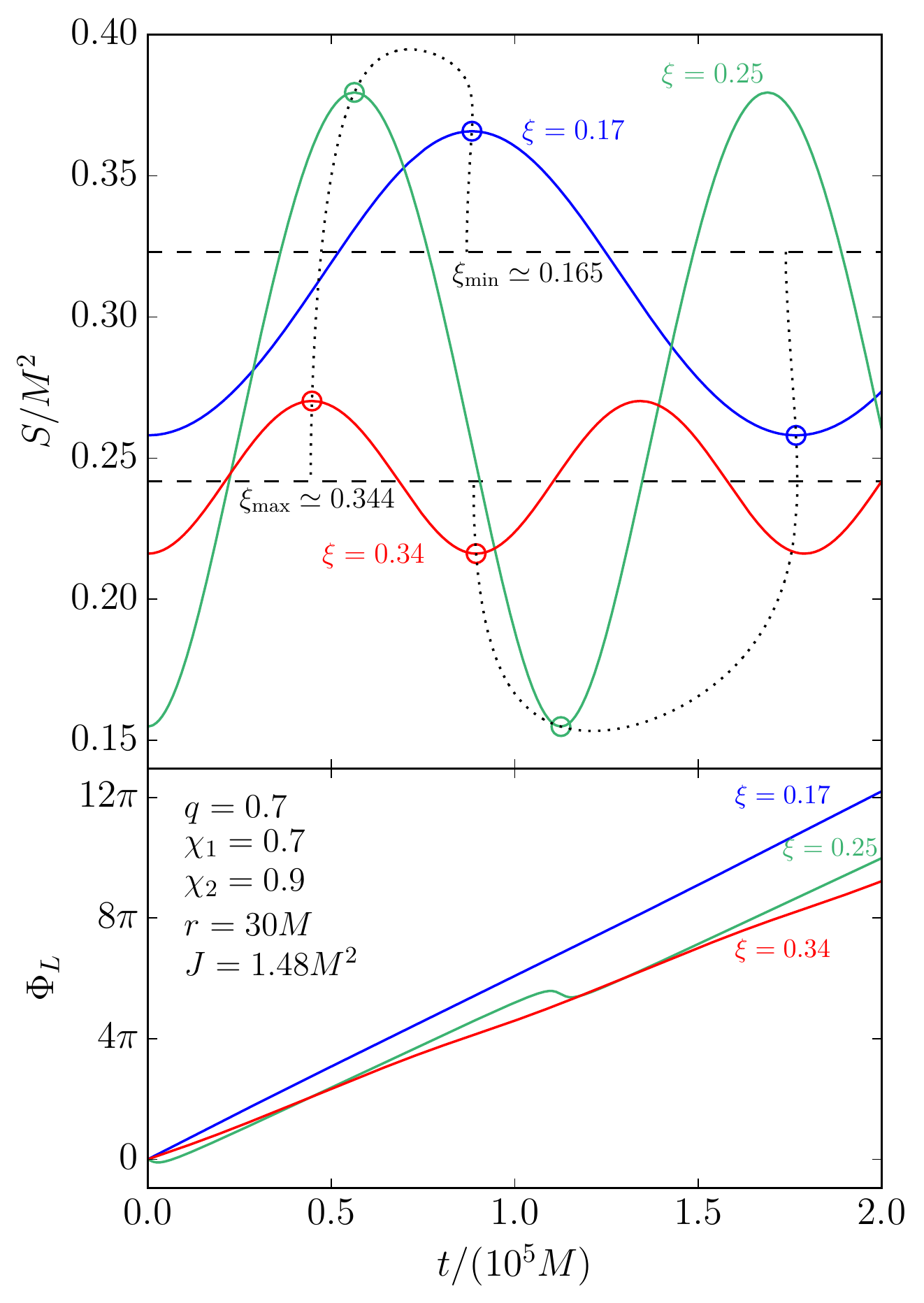}
\caption{Time-dependent solutions for the total-spin magnitude ${S}$ (top panel) and the orbital-angular-momentum
phase $\Phi_L$ (bottom panel).  We set $q=0.7$, $\chi_1=0.7$, $\chi_2=0.9$, $r=30M$ and $J=1.48 M^2$ and integrate
Eq.~(\ref{E:dSdt}) for three values of $\xi$ corresponding to the three different spin morphologies: $\Delta\Phi$ oscillates
about 0 ($\xi=0.17$, blue), circulates ($\xi=0.25$, green), and oscillates about $\pi$ ($\xi=0.34$, red).  Initial conditions
have been chosen such that  $S=S_-$ and $\Phi_L=0$ at $t=0$. The oscillations in $S$ induce small wiggles in $\Phi_L$
on top of a mostly linear drift. Spin-orbit resonances (horizontal dashed lines, top panel) correspond to configurations for
which $S$ is constant and can be interpreted as zero-amplitude limits of generic oscillatory solutions.  The projections of
the effective potentials, i.e. parametric curves $[\tau(\xi)/2, S_+(\xi)]$ and $[\tau(\xi), S_-(\xi)]$, are shown with dotted lines.
An animated version of this figure is available online~\cite{DGwebsite}.} \label{S_time}
\end{center}
\end{figure}

Although $S$ fully parametrizes the precessional dynamics, time-dependent expressions may be useful as well.  The BBH
spins obey the 2PN precession equations
\cite{1995PhRvD..52..821K,2008PhRvD..78d4021R,2006PhRvD..74j4033F,2006PhRvD..74j4034B} 
\begin{subequations}\label{preceq}
\begin{align}
\frac{d \mathbf{S_1}}{dt} = \frac{1}{2 r^3} \left[ (4+3q) \mathbf{L} - \frac{3q M^2 \xi  }{1+q} \mathbf{\hat L} + \mathbf{S_2}\right] \times \mathbf{S_1}\,,\\
\frac{d \mathbf{S_2}}{dt} = \frac{1}{2 r^3} \left[ \left(4+\frac{3}{q}\right) \mathbf{L} - \frac{3M^2 \xi}{1+q}  \mathbf{\hat L} + \mathbf{S_1}\right] \times \mathbf{S_2} \,,
\end{align}
\end{subequations}
which include the  quadrupole-monopole interaction computed in Ref.~\cite{2008PhRvD..78d4021R}. These equations
are averaged over the binary's orbital period $t_{\rm orb}$ and describe the evolution of the spins on the precession
timescale  $t_{\rm pre}$.  Equations~(\ref{preceq}) imply that the orbit-averaged evolution of $S=|\mathbf{S_1}+\mathbf{S_2}|$
is given by:
\begin{align} \label{E:dSdt}
\frac{dS}{dt} &= -\frac{3(1-q^2)}{2q} \frac{S_1S_2}{S} \frac{(\eta^2M^3)^3}{L^5} \left( 1 - \frac{\eta M^2 \xi}{L}  \right)
\notag \\
& \times \sin\theta_1 \sin\theta_2 \sin\Delta\Phi \,.
\end{align}
Integrating Eq.~(\ref{E:dSdt}) yields solutions $S(t)$, and that specifies $\mathbf{L}$, $\mathbf{S}_1$, and $\mathbf{S}_2$ as
functions of time through substitution into Eqs.~(\ref{momenta}).  Some examples of $S(t)$ for different values of $\xi$ are
shown in the top panel of Fig.~\ref{S_time}.

These time-dependent solutions confirm the scenario outlined in Sec.~\ref{subseceffpot}, with $S$ oscillating between
two turning points $S_-$ and $S_+$ at which $dS/dt=0$.  At these turning points, the three angular momenta are coplanar
[from Eq.~(\ref{E:dSdt}), $dS/dt=0$ implies either $\sin\Delta\Phi=0$  or  $\sin\theta_i=0$]
and the BBHs lie on the effective potentials 
\mbox{($\xi_\pm(S_\pm)=\xi$)}.   The spin-orbit resonances $\xi_{\rm min}$ and $\xi_{\rm max}$ are shown with dashed lines in
Fig.~\ref{S_time} and correspond to the zero-amplitude limits of the generic  oscillatory solutions.  From Eq.~(\ref{E:dSdt}),
we can define the precessional period $\tau$ as the time needed to complete a full cycle in $S$,
\begin{align}
\tau(L, J, \xi) = 2 \int_{S_-}^{S_+}\frac{dS}{|dS/dt|}\,.
\end{align}
The precession timescale $t_{\rm pre} \sim (2\pi M/\eta)(r/M)^{5/2}$ provides an order-of-magnitude estimate for this exact
precessional period.  The period $\tau$ remains finite at the spin-orbit resonances $\xi_{\rm min}$ and $\xi_{\rm max}$
in much the same way that the period of a simple harmonic oscillator remains finite in the limit of small oscillations.

The time evolution of the three angular momenta $\mathbf{L}$, $\mathbf{S_1}$ and $\mathbf{S_2}$ is fully given by 
Eqs.~(\ref{momenta}) and (\ref{E:dSdt}) when described in the non-inertial frames of Fig.~\ref{twoframes}. However,
$\mathbf J$ and $\mathbf L$ will generally not be confined to a plane in an inertial frame. The direction of $\mathbf J$ is
fixed on the precession time scale $t_{\rm pre}$, and hence so is $\mathbf{\hat z}$. The two remaining basis vectors will
precess about the $z$-axis  
\begin{equation}
\label{dxdtdydt}
\frac{d\hat{\mathbf{x}}}{dt} = \Omega_z \hat{\mathbf{z}} \times \hat{\mathbf{x}} = \Omega_z \hat{\mathbf{y}}\,, \quad
\frac{d\hat{\mathbf{y}}}{dt} = \Omega_z \hat{\mathbf{z}} \times \hat{\mathbf{y}} = -\Omega_z \hat{\mathbf{x}}\,.
\end{equation}
The solution to these two equations gives $\hat{\mathbf{x}}(t)$ and $\hat{\mathbf{y}}(t)$ and hence $\mathbf{L}(t)$,
$\mathbf{S}_1(t)$, and $\mathbf{S}_2(t)$ in an inertial frame from  Eqs.~(\ref{momenta}) and (\ref{xiSvaphi}).  The orbital
angular momentum $\mathbf{L}$ precesses about $\mathbf{J}$ with  frequency $\Omega_z$ given by
\cite{2015PhRvL.114h1103K}
\begin{align} \label{E:Omegaz}
\begin{aligned}
\Omega_z 
&=\frac{J}{2} \left( \frac{\eta^2M^3}{L^2} \right)^3 \bigg\{ 1 + \frac{3}{2\eta} \left( 1 - \frac{\eta M^2 \xi}{L} \right)
 \\ 
& -\frac{3(1+q)}{2qA_1^2A_2^2} \left(1 - \frac{\eta M^2 \xi}{L} \right)[4(1-q)L^2(S_1^2 - S_2^2)
 \\
& -(1+q)(J^2 - L^2 -S^2)(J^2 - L^2 -S^2 - 4\eta M^2L\xi)] \bigg\}\,.
\end{aligned}
\end{align}
This equation can be derived by substituting Eqs.~(\ref{preceq}) and (\ref{dxdtdydt}) into the time derivative of
Eq.~(\ref{bigSvector}).  For concreteness, let us specify an inertial frame such that $\mathbf{L}$ lies in the $xz$-plane at
$S=S_-$.  At the point on a precession cycle specified by the total-spin magnitude $S$, the direction of $\mathbf{L}$ is 
specified by the polar angles $\theta_L$ from Eq.~(\ref{E:cstL}) and the azimuthal angle 
\begin{align}
\Phi_L = 
\begin{dcases}
\int_{S_-}^{S}  \Omega_z \frac{dS}{|dS/dt|} 
&{\rm for} \quad S:S_- \to S_+\\
\frac{\alpha}{2} +\int_{S}^{S_+}  \Omega_z \frac{dS}{|dS/dt|}  
&{\rm for} \quad S:S_+ \to S_-\\
\end{dcases}
\end{align} 
where the two cases refer to the first and the second half of the precession cycle, and
\begin{align}
\label{eqalpha}
\alpha(L, J, \xi) = 2 \int_{S_-}^{S_+}  \Omega_z \frac{dS}{|dS/dt|}
\end{align}
is the total change in the azimuthal angle $\Phi_L$ over a full precession cycle.
Solutions $\Phi_L(t)$ are shown in the bottom panel of Fig.~\ref{S_time}. The angle $\Phi_L$ mainly exhibits a linear drift
due to the leading PN order term in Eq.~(\ref{preceq}).  Spin-spin couplings are of higher PN order and cause small
wiggles on top of this linear drift.  Binaries in spin-orbit resonances ($\xi_{\rm  min}$ and $\xi_{\rm max}$) precess at a 
constant rate $\Omega_z$ with all three vectors $\mathbf{L}$, $\mathbf{S_1}$, and $\mathbf{S_2}$ jointly precessing about
$\mathbf{J}$.  Just as $\Delta\Phi$ is ill defined if either of the $\mathbf{S_i}$ is aligned with $\mathbf{L}$ ($\cos\theta_i =
\pm1$), $\Phi_L$ and thus $\alpha$ is ill defined if $\mathbf{L}$ is aligned with $\mathbf{J}$ ($\cos\theta_L = \pm1$).  This
occurs for values of $J$ and $\xi$ for which $S_-=S_{\rm min}=|J-L|$ or $S_+=S_{\rm max}=J+L$, corresponding to some
of the transitions between the different classes of the evolution of $\varphi'$ (dashed lines in Fig.~\ref{varphi_evolution}).

We stress here that the time-dependent expressions reported in this section are  only valid on times
$t\sim \tau \ll t_{\rm RR}$, i.e. when the precessional dynamics approximately decouples from the inspiral.  This
approximation breaks down at small separations, where the difference between the three timescales is smaller
(cf. Sec.~\ref{transferbin}).

\section{Precession-averaged evolution on the inspiral timescale}
\label{sec:insptime}

The previous section focused on spin dynamics on the precessional timescale.  We now consider how spin precession
evolves as BBHs inspiral due to radiation reaction.  Our main tool is a precession-averaged equation to model the binary
inspiral (derived in Sec.~\ref{averaging} below) that will allow us to overcome numerical limitations of our previous
analyses~\cite{2010PhRvD..81h4054K, 2010ApJ...715.1006K, 2012PhRvD..85l4049B, 2013PhRvD..87j4028G,
2014PhRvD..89l4025G} and evolve BBHs inwards from arbitrarily large separations (Sec.~\ref{largeseplim}). This
improved computational scheme relying on our new multi-scale analysis allows us to more efficiently ``transfer" BBHs from
the large separations where they form astrophysically down to the small separations relevant for GW detection. In
Sec.~\ref{transferbin} we compare the results of our precession-averaged evolution against  the standard integration of the
merely orbit-averaged spin-precession equations.

\subsection{Averaging the average}
\label{averaging}

In the usual PN formulation (see e.g. Ref.~\cite{1995PhRvD..52..821K}), the timescale hierarchy $t_{\rm orb} \ll t_{\rm pre} \ll
t_{\rm RR}$ is exploited to average the evolution equations for $\mathbf{L}$, $\mathbf{S}_1$, and $\mathbf{S}_2$ over the
orbital period $T$.  We already saw above how this orbit averaging can be used to increase the computational efficiency
with which spin precession can be calculated [Eq.~(\ref{preceq}) can be integrated with time steps $t_{\rm orb} \ll \Delta t
\ll t_{\rm pre}$ much longer than the orbital timescale].  Radiation reaction can be similarly orbit averaged:
\begin{equation} \label{E:dvecJdt_orb}
\left\langle \frac{d\mathbf{L}_{\rm RR}}{dt} \right\rangle_{\rm orb} =\frac{1}{T} \int_0^{2\pi} \frac{d\mathbf{L}_{\rm RR}}{dt}\,\,
\frac{d \psi}{d \psi/dt}~,
\end{equation}
where ${d\mathbf{L}_{\rm RR}}/{dt}$ is the instantaneous change in the orbital angular momentum due to GW radiation
reaction and $\psi$ is the true anomaly parametrizing the orbital motion.  The flux ${d\mathbf{L}_{\rm RR}}/{dt}$ depends
implicitly on both $\psi$ and the angular momenta $\mathbf{L}$, $\mathbf{S}_1$, and $\mathbf{S}_2$; the former
dependence can be averaged over since we have analytic solutions to the orbital motion as function of $\psi$, while the
angular momenta may be held fixed, since they barely evolve over an orbital period.  Spin precession may be calculated on
the radiation-reaction timescale by numerically integrating the coupled system of ordinary differential equations (ODEs)
given by Eqs.~(\ref{preceq}) and (\ref{E:dvecJdt_orb}) with the time step $\Delta t$ given above.

We derived analytic solutions to the orbit-averaged spin-precession equations (\ref{preceq}) in Sec.~\ref{sec:prectime}
that depend on the magnitudes $L$ and $J$ that evolve on the radiation-reaction timescale $t_{\rm RR}$.  In a similar
spirit to the orbit averaging discussed above, we can use these solutions to {\it precession average} the evolution equations
for $L$ and $J$.  We define the precession average of some scalar quantity $X$ to be
\begin{equation} \label{precaverageX}
\left\langle X \right\rangle_{\rm pre} 	\equiv \frac{2}{\tau}
\int_{S_-}^{S_+} \langle X \rangle_{\rm orb} \frac{dS}{|dS/dt|} 
\end{equation}
where $dS/dt$ is given as a function of $S$ in Eq.~(\ref{E:dSdt}). 
We can hold $L$, $J$, and $\xi$ fixed on the right-hand side of this equation because they barely evolve during a
precession cycle, much as we held the vectorial angular momenta fixed in the orbit averaging since they evolve on the 
longer timescale $t_{\rm pre} \gg t_{\rm orb}$. 

Since $\xi$ is conserved by radiation reaction at 2.5PN order \cite{2001PhRvD..64l4013D,2008PhRvD..78d4021R}, we
need only find precession-averaged evolution equations for $L$ and $J$ to evolve our spin-precession solutions on the
radiation-reaction timescale $t_{\rm RR}$.  Since $L^2 = \mathbf{L} \cdot \mathbf{L}$, $dL/dt = {\mathbf{\hat L}} \cdot
d\mathbf{L}_{\rm RR}/dt$ and the precession-averaged evolution of $L$ is given by 
\begin{align}
\label{E:dLdt_pre}
\begin{aligned}
\left\langle \frac{dL}{dt} \right\rangle_{\rm pre}  &= \frac{2}{\tau} \int_{S_-}^{S_+} {\mathbf{\hat L}} \cdot \left\langle   \frac{d\mathbf{L}_{\rm RR}}{dt}\right\rangle_{\rm orb} \frac{dS}{|dS/dt|}~.
\end{aligned}
\end{align}
We similarly have $dJ/dt = {\mathbf{\hat J}} \cdot d\mathbf{J}_{\rm RR}/dt$, but since $\mathbf{J} = \mathbf{L} + \mathbf{S}_1
+  \mathbf{S}_2$ and GW emission does not directly affect the individual spins ($d\mathbf{S}_{i,{\rm RR}}/dt = 0$),
$d\mathbf{J}_{\rm RR}/dt = d\mathbf{L}_{\rm RR}/dt$ and we have
\begin{align} \label{E:dJdt_pre1}
\begin{aligned}
\left\langle \frac{dJ}{dt} \right\rangle_{\rm pre} &= \frac{2}{\tau} \int_{S_-}^{S_+}
\mathbf{\hat J} \cdot \left\langle   \frac{d\mathbf{L}_{\rm RR}}{dt}\right\rangle_{\rm orb}  \frac{dS}{|dS/dt|} \,.
\end{aligned}
\end{align}
The orbit-averaged angular momentum flux $\langle{d\mathbf{L}_{\rm RR}}/{dt}\rangle_{\rm orb}$ up to 
1PN is given by
\cite{1995PhRvD..52..821K}
\begin{align} 
\label{E:PNLflux}
&\left\langle \frac{d\mathbf{L}_{\rm RR}}{dt} \right\rangle_{\rm orb} = -\frac{32}{5} \frac{\eta L}{M} \left( \frac{M}{r} \right)^4 \notag \\
 &\quad \times\left[\left( 1 - \frac{2423 + 588\eta}{336} \frac{M}{r} \right)
 \hat{\mathbf{L}} 
 + \mathcal{O}\left(\frac{M}{r}\right)^{{3/2}}\right]~.
\end{align}
Note that this expression is parallel to $\mathbf{\hat L}$ and independent of $S$.  Substituting this result into
Eq.~(\ref{E:dJdt_pre1}) yields
\begin{align} \label{E:dJdt_pre2}
\begin{aligned}
\left\langle \frac{dJ}{dt} \right\rangle_{\rm pre} =\frac{2}{\tau} \int_{S_-}^{S_+} 
\mathbf{\hat L} \cdot
\left\langle \frac{d\mathbf{L}_{\rm RR}}{dt}\right\rangle_{\rm orb}
 \cos\theta_L \frac{dS}{|dS/dt|}~,
\end{aligned}
\end{align}
where we used Eq.~\eqref{JLeq}, and $\cos\theta_L$ is given in Eq.~(\ref{E:cstL}) as a function of $S$.  Finally, Eqs.~(\ref{E:dLdt_pre}) and (\ref{E:dJdt_pre2}) together lead to 
\begin{align} \label{ourODE}
\begin{aligned}
\left\langle \frac{dJ}{dL} \right\rangle_{\rm pre} 
= \frac{1}{2LJ} (J^2 + L^2 - \langle S^2 \rangle_{\rm pre})~,
\end{aligned}
\end{align}
which reduces the computation of BBH spin precession on the radiation-reaction timescale to solving a single ODE
\cite{2015PhRvL.114h1103K}!  Equation~(\ref{ourODE}) is independent of the details of spin precession (which are encoded in
$\langle S^2 \rangle_{\rm pre}$) and is also independent of the PN expansion for
$\langle {d\mathbf{L}_{\rm RR}}/{dt}\rangle_{\rm orb}$ provided this is parallel to $\mathbf{\hat L}$ and independent of $S$. 
As shown in Eq.~(\ref{E:PNLflux}), both of these conditions are satisfied at 1PN level but break down at higher PN order. 
We address the range of validity of our approach in Sec.~\ref{transferbin}, where we also perform extensive comparisons
with full integrations of the conventional orbit-averaged equations.

\begin{figure}
\begin{center}
\includegraphics[width=0.9\columnwidth]{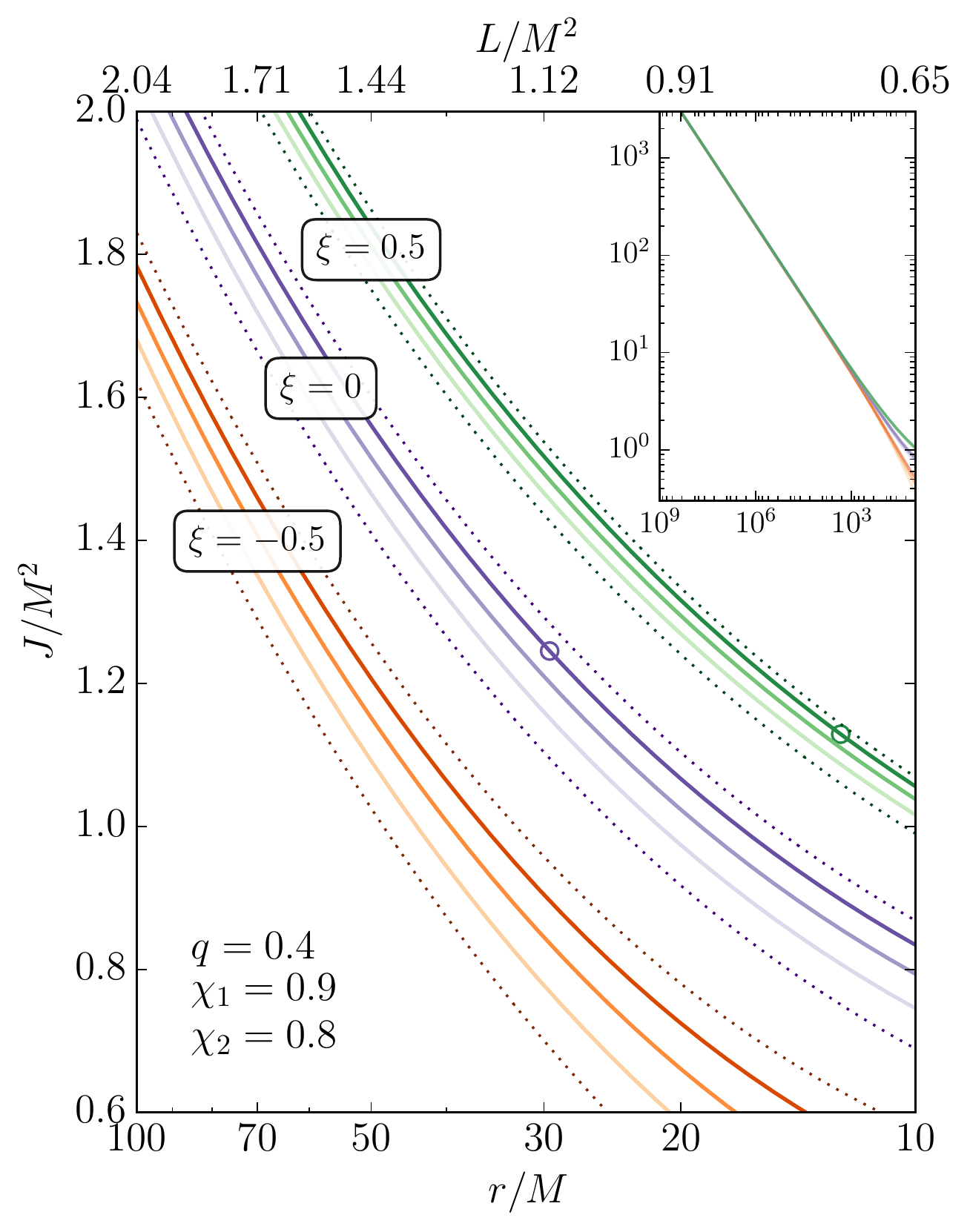}
\caption{Evolution of the total angular momentum magnitude $J$ during the inspiral. Three binary configurations are
considered here: $\xi=-0.5$ (orange), $0$ (purple) and $0.5$ (green) for  $q=0.4$, $\chi_1=0.9$, $\chi_2=0.8$.
Equation~(\ref{ourODE}) is solved for several different initial conditions (solid lines, sequential colors) as the separation $r$ and 
the angular momentum $L=\eta (r M^3)^{1/2}$ decrease. Solutions are bounded at all separations by the spin-orbit
resonances (dotted lines) which extremize the allowed value of $J$ for fixed $\xi$. Two of the binaries pictured here cross
one of the resonant conditions $\alpha=2\pi n$ (empty circles) where changes in the direction ${\mathbf{\hat J}}$ are
expected. The inset shows the same evolutions for a wider separation range.} \label{Jflow}
\end{center}
\end{figure}

Examples of solutions to Eq.~(\ref{ourODE}) are shown in Fig.~\ref{Jflow}, where $J$ is evolved from $r=10^9 M $ to
$r=10M$.  Solutions $J(r)$ are bounded at all separations by the spin-orbit resonances $\xi_{\rm min}$ and $\xi_{\rm max}$
which extremize the magnitude $J$ for each fixed $\xi$ (cf. Sec.~\ref{subsecmorph} and Fig.~\ref{threeJxi}).  We perform
ODE integrations using the \textsc{lsoda} algorithm \cite{hindmarsh1982odepack} as wrapped by the \textsc{python} module
\textsc{scipy} \cite{Jones:2001aa}; integrations of Eq.~(\ref{ourODE}) are numerically feasible for arbitrary values of $q<1$,
$\chi_1\leq 1$, $\chi_2\leq 1$,
and \emph{arbitrarily} large initial separation. 

Our solutions to the spin-precession equations also depend on the direction ${\mathbf{\hat J}}$, since this defines the
$z$-axis in the orthonormal frame of Fig.~\ref{twoframes}.  The precession-averaged evolution of this direction is
\begin{align} \label{E:dJhatdt_pre}
\begin{aligned}
\left\langle \frac{d\mathbf{\hat{J}}}{dt} \right\rangle_{\rm pre}  &= \frac{1}{J} \left\langle \left\langle
\frac{d\mathbf{L}_{\rm RR}}{dt} \right\rangle_{\rm orb} -
\frac{dJ}{dt} \mathbf{\hat{J}} \right\rangle_{\rm pre}
\end{aligned}
\end{align}
which is proportional to the precession average of the total angular momentum radiated perpendicular to
${\mathbf{\hat J}}$.  Although the vector given by the right-hand side of Eq.~(\ref{E:dJhatdt_pre}) will generally not vanish
over a single precession cycle, if the angle $\alpha$ given by Eq.~(\ref{eqalpha}) above is not an integer multiple of $2\pi$
this vector will precess about ${\mathbf{\hat J}}$ in an inertial frame.  This implies that ${\mathbf{\hat J}}$ will precess in a
narrow cone in an inertial frame on the radiation-reaction timescale remaining approximately constant
\cite{1994PhRvD..49.6274A,2009ApJ...704L..40B}.   As shown for some of the binaries of Fig.~\ref{Jflow}, the condition
$\alpha = 2\pi n$ for integer $n$ is indeed satisfied in generic inspirals at meaningful separations.  Preliminary results
indicate that interesting spin dynamics arises at these newly identified resonances \cite{ZhaoPrep}.  In this paper, we
restrict our attention to the relative orientations of the three angular momenta as specified by the three angles in
Eqs.~(\ref{t1t2dphisolutions}).
  
\subsection{The large-separation limit}
 \label{largeseplim}
 
We can gain additional physical insight by examining Eq.~(\ref{ourODE}) in the large-separation limit $L/M^2\to \infty$.  Let
us define 
\begin{align}\label{kappadef}
\kappa\equiv \frac{J^2 - L^2}{2 L}\,,
\end{align}
such that Eq.~(\ref{ourODE}) becomes
\begin{align} \label{E:dKdL}
\frac{d \kappa}{dL} = - \frac{\langle S^2 \rangle_{\rm pre}}{2 L^2}\,.
\end{align}
The right-hand side vanishes at large separations where $S\ll L$, implying that 
\begin{align}
\kappa_\infty \equiv \lim_{r/M\to \infty} \kappa
\end{align}
is constant.  This implies that $\kappa$ provides a more convenient label for precessing BBHs at large separations
because it asymptotes to a constant while $J$ diverges.  At large separations $J$ evolves as 
\begin{align}
\label{Jlarge}
J = \sqrt{L(2\kappa + L)}  \simeq \sqrt{L\left( 2 \kappa_\infty + L\right)}\,,
\end{align}
 as  illustrated in the inset of Fig.~\ref{Jflow}. From Eq.~(\ref{kappadef}) and $\mathbf{J}=\mathbf{L}+\mathbf{S}$ one also obtains 
\begin{align} \label{kappaproj}
\kappa_\infty = \lim_{r/M\to \infty} \mathbf{S}\cdot \mathbf{\hat L}
\end{align}
implying that $\kappa$ asymptotes to the projection of the total spin onto the orbital angular momentum.  The constant $\kappa_\infty$
can be calculated for a binary at finite separation by integrating $d\kappa/dL$ all the way to $r/M\to\infty$.  This
integration can be performed by defining $u=1/2L$ such that $d\kappa/du = \langle S^2 \rangle_{\rm pre}$ can be
integrated over a compact domain. 

The two constants $\kappa_\infty$ and $\xi$ are linear combinations of the asymptotic values of the inner products
$\mathbf{\hat S_i} \cdot \mathbf{\hat L}$ defined in Eqs.~(\ref{t1t2dphisolutions}) in the large-separation limit.  The
constancy of these inner products at large separations is also apparent from Eqs.~(\ref{preceq}), where the
$\mathbf{S_i}$ will precess about $\mathbf{L}$ when spin-orbit coupling dominates over spin-spin coupling.  From
Eqs.~(\ref{E:xidef}) and (\ref{kappaproj}) one finds
\begin{subequations} \label{E:csinfty}
\begin{align} 
\cos\theta_{1\infty}&\equiv\lim_{r/M \to \infty}  \mathbf{\hat S_1} \cdot \mathbf{\hat L} = \frac{- M^2\xi + \kappa_\infty(1+q^{-1})}{S_1(q^{-1}-q)}\,,
\\
\cos\theta_{2\infty}&\equiv\lim_{r/M \to \infty}  \mathbf{\hat S_2} \cdot \mathbf{\hat L} = \frac{M^2\xi - \kappa_\infty(1+q)}{S_2(q^{-1}-q)}\,.
\end{align}
\end{subequations}
The terms in Eqs.~(\ref{t1t2dphisolutions}) proportional to $S^2$ become increasingly significant at smaller separations
and induce oscillations in $\theta_{i}$ on the precession timescale, while the breakdown of the asymptotic approximation to
$J(L)$ given in Eq.~(\ref{Jlarge}) causes $J$ (and hence $\theta_i$) to deviate on the radiation-reaction timescale
for BBHs with different values of $\xi$, as seen in Fig.~\ref{Jflow}.  The constraints $|\cos\theta_{1\infty}|\leq1$ and
$|\cos\theta_{2\infty}|\leq 1$ define the physically allowed values of $\xi$ and $\kappa_\infty$.  These parameters, or
equivalently $\theta_{1\infty}$ and $\theta_{2\infty}$, can be used to
identify an entire BBH inspiral (as far as the relative orientation of
the angular momenta is concerned) without reference to a
particular separation or frequency, as typically done in GW applications \cite{2014PhRvD..89l4025G,2014PhRvD..89f4048O,
2015PhRvD..91d2003V,2013PhRvD..87b4004C,
2014PhRvD..90b4018F,2014PhRvD..89f4048O,2014PhRvD..89j2005O}. 

\subsection{Efficient binary transfer}
\label{transferbin}

Our new precession-averaged equation for $dJ/dL$~(\ref{ourODE}) can be used to efficiently ``transfer" BBHs from the
large separations at which they form astrophysically to the smaller separations at which the GWs they emit become
detectable.  This equation can be integrated with a time step $t_{\rm pre} \ll \Delta t' \ll t_{\rm RR}$ much longer than the
time step $t_{\rm orb} \ll \Delta t \ll t_{\rm pre}$ on which merely orbit-averaged equations must be integrated.  This
greater efficiency comes at the cost of no longer being able to keep track of the precessional phase, in much the same way
that orbit-averaged equations do not explicitly evolve the orbital phase.  This is not a major problem for
population-synthesis studies however, because evolution over a timescale $\Delta t'$ will randomize the precessional phase, as
described below.  If one needs to track the precessional phase below a certain separation (such as that corresponding to
the lowest detectable GW frequencies) one can randomly initialize the phase at this separation and then employ
orbit-averaged equations.  The following procedure explicitly outlines how to evolve the spin orientations of a population
of BBHs from large to small separations.

\begin{enumerate}

\item Given a sample of BBHs specified by values of $q$, $\chi_1$ and $\chi_2$, choose a distribution $p_i(\theta_1, \theta_2, \Delta\Phi)$ for the angles that describes the spin orientations at an
initial separation $r_i$.  This initial distribution is determined by the interactions between BHs and their astrophysical
environment that lead to binary formation (cf. Refs.~\cite{2013PhRvD..87j4028G, 2008ApJ...682..474B, 2010ApJ...719L..79F,
2000ApJ...541..319K} on stellar-mass BHs and Refs.~\cite{2010MNRAS.402..682D, 2015arXiv150306807,
2014ApJ...794..104S, 2007ApJ...661L.147B,2013ApJ...774...43M} on supermassive BBHs).

\item Rewrite this initial distribution as a distribution $p_i(J, \xi)$ using the relations
\begin{subequations}
\begin{align}
&\qquad\begin{aligned}
S &= [S_1^2 + S_2^2 + 2S_1S_2(\sin\theta_1\sin\theta_2\cos\Delta\Phi \\ 
&\qquad\qquad\qquad\quad\qquad+ \cos\theta_1\cos\theta_2)]^{1/2}~, \end{aligned}\\
&\qquad J = [L^2 +  S^2 + 2L (S_1 \cos\theta_1 + S_2 \cos\theta_2)]^{1/2}~,\\
&\qquad \xi = \frac{qS_1\cos\theta_1 + S_2\cos\theta_2}{\eta M^2(1+q)}~.
\end{align}
\end{subequations}

\item Evolve each member of the distribution $p_i(J, \xi)$ to a smaller separation $r_f$ using Eq.~(\ref{ourODE})
for $dJ/dL$ ($\xi$ remains constant).  This yields a final distribution $p_f(J, \xi)$. 

\item For each member of the distribution $p_f(J, \xi)$, create a distribution of values of $S$ in the range $S_{-}(J, \xi) \leq S
\leq S_{+}(J, \xi)$ weighted by $(dS/dt)^{-1}$ given by Eq.~(\ref{E:dSdt}).  BBHs spend less time at values of $S$ where
the ``velocity'' $dS/dt$ is large.  This yields a distribution $p_f(S, J, \xi)$.

\item Convert $p_f(S, J, \xi)$ into a distribution of final angles $p_f(\theta_1, \theta_2, \Delta\Phi)$ using
Eqs.~(\ref{t1t2dphisolutions}) and a randomly chosen sign for $\Delta\Phi$.  

\end{enumerate}

\begin{figure*}[p]
\includegraphics[width=\textwidth]{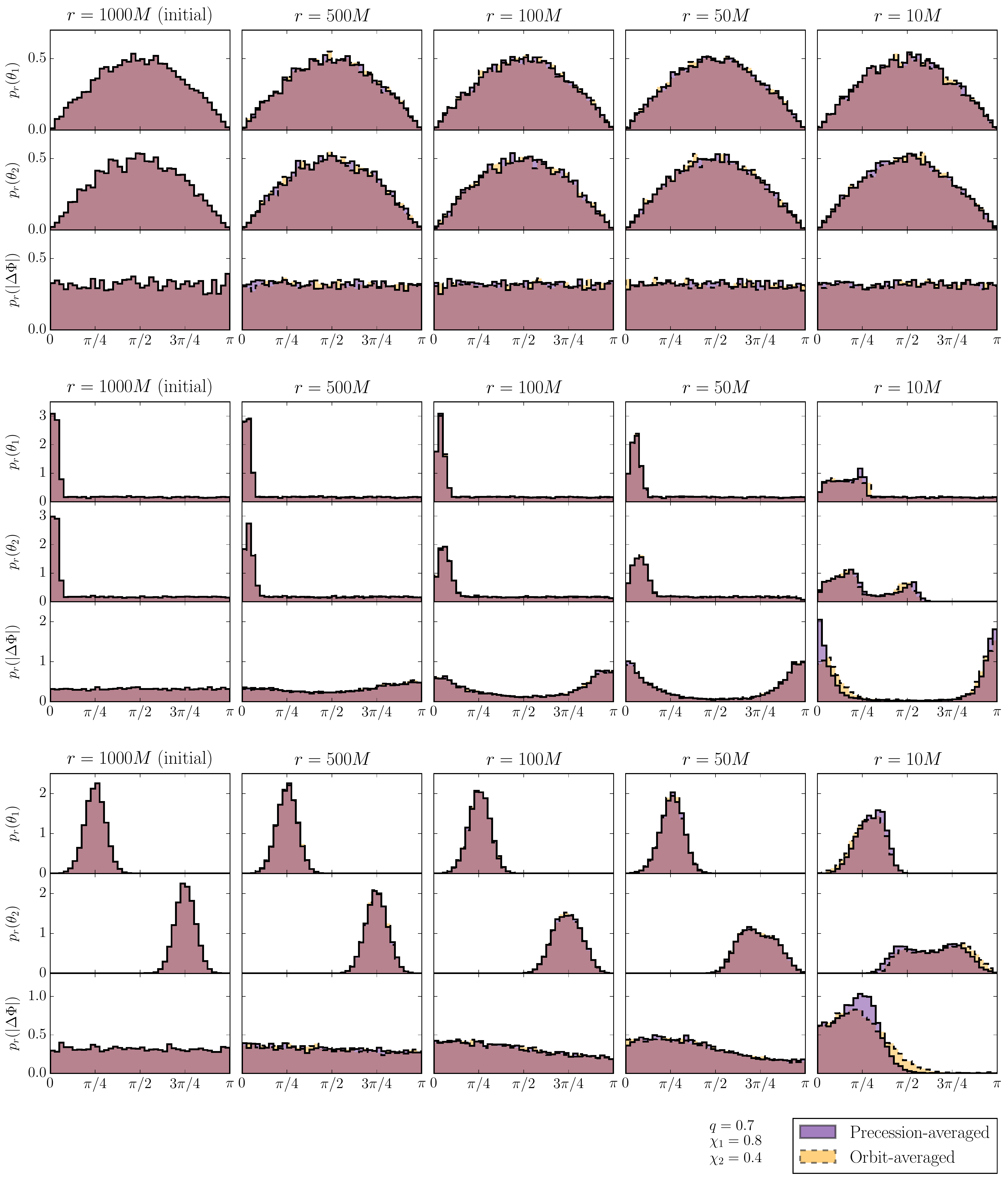}
\caption{Precession-averaged BBH inspirals as described in Sec.~\ref{transferbin} (purple/darker) compared to numerical
integration of the orbit-averaged PN equations \cite{2010PhRvD..81h4054K, 2012PhRvD..85l4049B} (orange/lighter).
Marginalized distributions of the spin angles $\theta_1$, $\theta_2$, and $|\Delta\Phi|$ (rows) are shown at several
separations along the inspirals [columns: $r_i=1000M$, $500M,100M,50M$, and $10M$].  The three initial spin
distributions are isotropic (top panels), one aligned BH (middle panels), and Gaussian spikes (bottom panels) as
described in Sec.~\ref{transferbin}.  The two approaches are in good agreement except for minor deviations in the
distribution of $\Delta\Phi$ at $r\sim 10M$.  We take $q=0.7$, $\chi_1=0.8$ and $\chi_2=0.4$ for all BBHs.
An animated version of this figure is available online~\cite{DGwebsite}.} \label{benchmark}
\end{figure*}

Examples of this binary transfer are given in Fig.~\ref{benchmark} for three different initial spin distributions.
\begin{enumerate}

\item {\it Isotropic sample} (top panels): Both spin vectors are isotropically distributed (flat uncorrelated distributions in
$\cos\theta_1$, $\cos\theta_2$ and $\Delta\Phi$).

\item {\it One aligned BH} (middle panels):  One BH spin (either the spin of the primary or the spin of the secondary) is aligned within $10^\circ$ of the orbital angular momentum,
while the other spin angle $\theta_i$ has a flat distribution in $[0^\circ,180^\circ]$; $\Delta\Phi$ is also flat in
$[-180^\circ,180^\circ]$.

\item {\it Gaussian spikes} (bottom panels): $\theta_1$ and $\theta_2$ have Gaussian distributions peaked at $45^\circ$
and $135^\circ$ with deviations of $10^\circ$; $\Delta\Phi$ is kept flat in $[-180^\circ,180^\circ]$.

\end{enumerate}
We evolve these distributions from $r_i=1000M$ to $r_f=10M$ and show marginalized distributions of the three angles
$\theta_1$, $\theta_2$, and $\Delta\Phi$ at several intermediate separations.  An animated version of this figure can be found online~\cite{DGwebsite}.  The isotropic sample remains isotropic, as found previously using the orbit-averaged
equations \cite{2007ApJ...661L.147B}.  A greater fraction of the BBHs in the distribution with one aligned BH undergo a
phase transition from a circulating to a librating morphology, as described in Sec.~\ref{sec:phtr} below and also found in
previous studies with the orbit-averaged equations \cite{2013PhRvD..87j4028G}.  If the angles $\theta_i$ initially have
Gaussian distributions, these Gaussians will spread out as the inspiral proceeds.

We use the BBH inspirals from $r_i=1000M$ to $r_f=10M$ shown in Fig.~\ref{benchmark} to compare the efficiency of our 
new precession-averaged approach to the integration of the standard -- i.e., orbit-averaged -- PN equations.
In the standard approach, one must numerically integrate ten coupled ODEs specifying the directions of the three angular
momenta and the magnitude of the orbital velocity; we use the PN equations quoted by
Refs.~\cite{2010PhRvD..81h4054K,2012PhRvD..85l4049B}.
We implement the same 2PN spin-precession equations\footnote{Higher-order PN corrections to the spin-precession equations have been computed in Refs.~\cite{2013CQGra..30e5007M,2013CQGra..30g5017B,2015arXiv150101529B}; their implementation is left to future work.} given by Eqs.~(\ref{preceq}) but include radiation reaction up to
3.5PN order, as in Eq.~(2.6) of Ref.~\cite{2010PhRvD..81h4054K}.  Integrations are performed using the same algorithm
specified above  \cite{Jones:2001aa,hindmarsh1982odepack}.
The agreement between the two approaches is seen to be excellent up to $r\sim 50 M$, and minor discrepancies emerge at smaller separations.

Two approximations made in the precession-averaged approach may explain these discrepancies.  While $\xi$ is held
constant throughout the inspiral in the precession-averaged approach (consistent with 2.5PN radiation reaction),
conservation of $\xi$ is not enforced in the orbit-averaged approach, which employs 3.5 PN radiation reaction.  The largest
deviations $\Delta\xi$ in the latter approach are of the order $10^{-10}$; $\xi$ is effectively constant in the PN regime
($r \gtrsim 10 M$).  Numerical-relativity simulations may be used to test conservation of $\xi$ at smaller separations.
We have verified that additional PN corrections in
  Eq.~(\ref{E:PNLflux}), implemented in our orbit-average code up to
  3.5PN, introduce very mild corrections to the evolution of $J$: the
  largest variations observed in our evolutions are of order  $\Delta J \sim10^{-2}$.

The second and less reliable approximation involves the timescale hierarchy itself.  The precession time $t_{\rm pre}\sim
(r/M)^{5/2}$ and radiation-reaction time $t_{\rm RR}\sim (r/M)^4$ become more comparable at lower separations, reducing
the effectiveness of our quasi-adiabatic approach.  The precession-averaging procedure defined in
Eq.~(\ref{precaverageX}) assumes that quantities like $L$ and $J$ varying on $t_{\rm RR}$ remain constant over a full
precession cycle $\tau$, but this assumption will break down as the timescale hierarchy becomes invalid.

Figure~\ref{benchmark} shows that differences between the two approaches are most pronounced in $p_r(\Delta\Phi)$.  This
variable is the most sensitive to the precessional dynamics; predictions for the angles $\theta_1$ and $\theta_2$ remain
reasonably accurate even at $r\sim 10M$.  The differences seem to average out for wider distributions (top panels) but
become more evident for more compact initial distributions (bottom panels).  Averaging over the precessional dynamics
prevents us from tracking the precession phase, implying that the two approaches will make different predictions for
quantities (like $S$ and $\Delta\Phi$) varying on the precession timescale when the initial separation is sufficiently small
that memory of the initial phases has not been fully forgotten.  Predictions of physical quantities varying on the
radiation-reaction timescale (like $J$ and the precession morphology) will remain robust down to small separations, as
explored in Secs.~\ref{msasymmetry} and \ref{binmemory} below.

\begin{figure}
\includegraphics[width=\columnwidth]{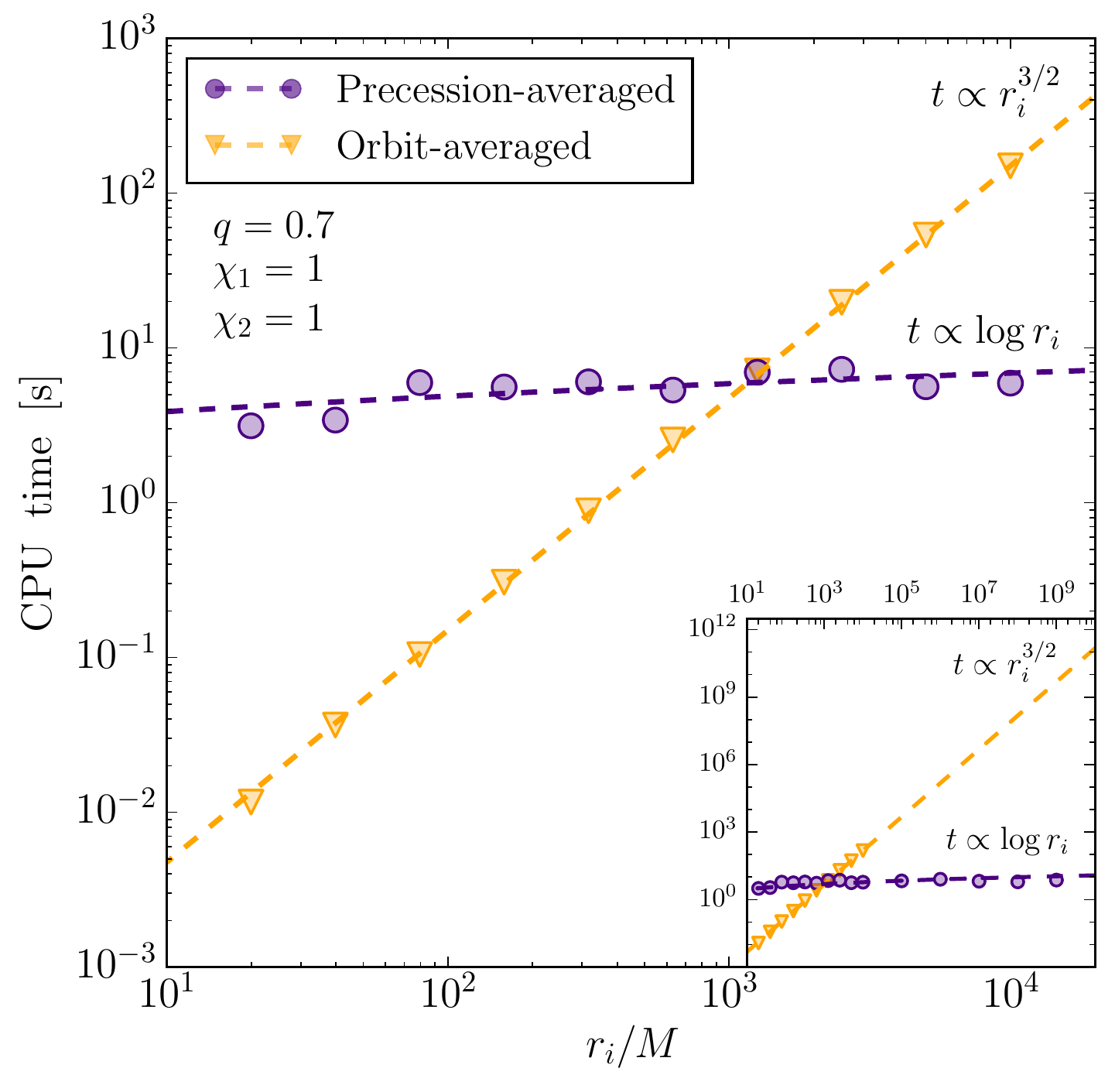}
\caption{CPU time needed to evolve BBHs from an initial separation $r_i$ to a final separation $r_f=10M$ using
our new precession-averaged approach (purple circles) and the standard orbit-averaged approach (orange triangles).
Each CPU time is averaged over $N=100$ executions with isotropic initial spin orientation (flat distributions in
$\cos\theta_{1}$, $\cos\theta_{2}$ and $\Delta\Phi$).  Dashed lines show the expected scalings: $t\propto r_i^{3/2}$ for the
orbit-averaged approach and $t\propto \log r_i$ for our new precession-averaged approach.  These computations have
been performed on a single core of a 2013 Intel i5-3470 3.20GHz CPU.} \label{timing}
\end{figure}

We compare the computational efficiency of the precession- and orbit-averaged approaches in Fig.~\ref{timing}.   Isotropic
samples of 100 BBHs are transferred from large initial separations $r_i$ to a final separation $r_f=10M$.  The CPU time
required by the two approaches scales differently with the initial separation.  The orbit-averaged (OA) equations must be
integrated with a time step shorter than the precession time, implying that the total number of time steps scales as 
\begin{align}
N_{\rm OA}\propto \int_{r_{f}}^{r_{i}} \frac{dr}{ \dot r_{GW} \;t_{\rm pre}}\sim r_i^{3/2}\,,
\end{align}
where $\dot r_{GW}\propto r^{-3}$ as given by the quadrupole formula \cite{1963PhRv..131..435P,1964PhRv..136.1224P}. 
The ratio $t_{\rm RR}/t_{\rm pre} \propto r^{3/2}$ increases dramatically at large separations leading to a corresponding
increase in the computational cost.  In the precession-averaged (PA) approach, the integration of $dJ/dL$ in
Eq.~(\ref{ourODE}) only requires time steps proportional to $L$, hence
\begin{align}
N_{\rm PA} \propto \int^{L_i}_{L_f} \frac{dL}{L}  \sim \log(L_i) \propto \log(r_i)\,.
\end{align}
The precession-averaged approach is very efficient at large separations because the solutions to Eq.~(\ref{ourODE})
become very smooth in this limit as seen from Eq.~(\ref{Jlarge}) and Fig.~\ref{Jflow}.  Precession-averaged inspirals may
even be computed from infinite separations through a change of variables to $u \equiv (2L)^{-1}$.
The integrator spends most of the computational time at small separations, where spin effects -- notably the numerical
evaluation of $S_\pm$ -- need to be tracked with high accuracy to avoid violations of the constraints~(\ref{Jlim}).  As shown
in Fig.~\ref{timing}, these expected scalings are well reproduced by both of our codes.

In addition to the time needed to integrate Eq.~(\ref{ourODE}), the precession-averaged approach must generate a final
distribution for $S$ (step 4 above), implying that the computational cost does not go to zero as $r_i\to r_f$.  While this step
makes the calculation of a single BBH inspiral non-deterministic and more expensive, precession averaging effectively
reduces the dimensionality of the BBH population during the inspiral.  If the $n$ members of this final distribution for $S$
are regarded as distinct binaries, the total number of integrations required to produce a fixed number of BBHs at $r_f$ is
reduced by a factor of $n$ in the precession-averaged approach compared to the orbit-averaged approach.

\begin{figure*}
\includegraphics[width=\textwidth]{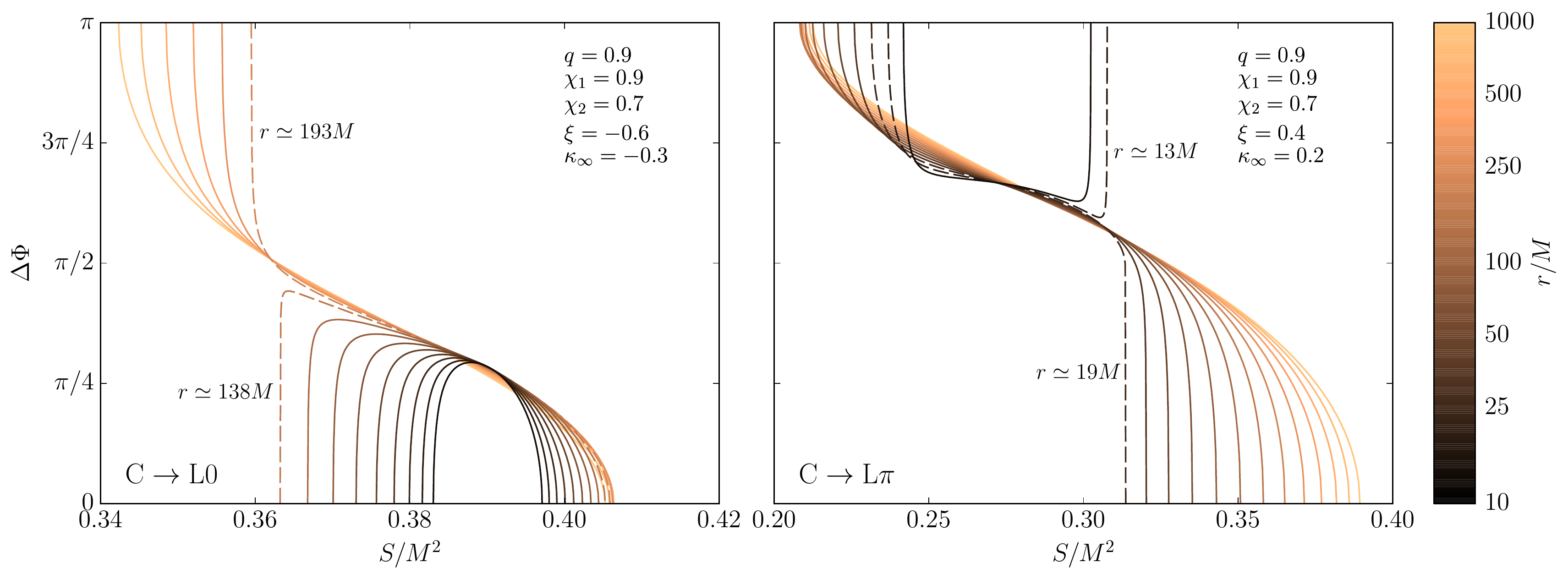}
\caption{Precessional solutions $\Delta\Phi(S)$ of Eqs.~(\ref{t1t2dphisolutions}) as $J$ and $L$ evolve during inspirals
according to Eq.~(\ref{ourODE}).  These solutions are colored according to the separation $r/M$ as shown in the color bar
on the right (orange/lighter for large separations and black/darker for small separations).  Binaries in the left (right) panel
transition from the circulating morphology to the morphology in which $\Delta\Phi$ librates about 0 ($\pi$) at the transition
radius $r_{\rm tr} \simeq 152 M$ ($18.9M$); separations bracketing the transition radius are marked with dashed lines.  Parameters are set to the values indicated in the legends.  An animated version of this figure is available online~\cite{DGwebsite}.}
\label{discontinuity}
\end{figure*}

\section{Morphological phase transitions}
\label{sec:phtr}

As BBHs inspiral on the radiation-reaction timescale, they can transition between the spin-precession morphologies
described in Sec.~\ref{subsecmorph}.  BBH spins predominantly circulate at large separations but increasingly transition
into one of the two librating morphologies as spin-spin coupling becomes important (Sec.~\ref{phasetransitionssec}).
The probability of encountering one of these morphological phase transitions during the inspiral depends on the
asymmetry between the masses and the spin magnitudes of the two BBHs (Sec.~\ref{msasymmetry}).  Asymmetric binaries
are more likely to circulate, while BBHs with comparable mass and spin ratios populate the librating morphologies.  BBH
spin morphologies at finite separations can be determined from their asymptotic spin orientations $\cos\theta_{i\infty}$ (or
equivalently $\xi$ and $\kappa_\infty$) as discussed in Sec.~\ref{binmemory}.

\subsection{Phenomenology of phase transitions} 
\label{phasetransitionssec}

As extensively discussed in Sec.~\ref{subsecmorph}, BBH spin precession can be unambiguously classified into one of
three morphologies depending on the values of $q$, $\chi_1$, $\chi_2$, $\xi$, $r$ (or $L$), and $J$.
While the first four of these parameters remain constant throughout the inspiral, $r$ and $J$ evolve on the
radiation-reaction timescale according to Eq.~(\ref{ourODE}).  Binaries may therefore change their precessional
morphology while evolving towards merger.  The boundaries between different morphologies (cf. Sec.~\ref{subsecmorph})
are set by the (anti)alignment condition $\sin\theta_i=0$; the binary morphology changes whenever radiation reaction
brings $J$ and $L$ to values that satisfy this condition (which can only occur on the effective-potential loop $\xi_\pm(S)$, as
seen in Fig.~\ref{threeeffpot}).  Figure \ref{discontinuity} shows two examples of these phase transitions.  At the radii
$r_{\rm tr}$ where phase transitions occur, $\Delta\Phi$ changes discontinuously either at $S_-$ (left panel) or $S_+$
(right panel), causing the solutions $\Delta\Phi(S)$ of Eqs.~(\ref{t1t2dphisolutions}) to transition between the qualitatively
different shapes seen in the bottom panel of Fig.~\ref{angle_solutions}.  The BBHs in the left (right) panel evolve from the
circulating morphology to the morphology in which $\Delta\Phi$ oscillates about 0 ($\pi$).

\begin{figure*}[p]
\includegraphics[width=\textwidth]{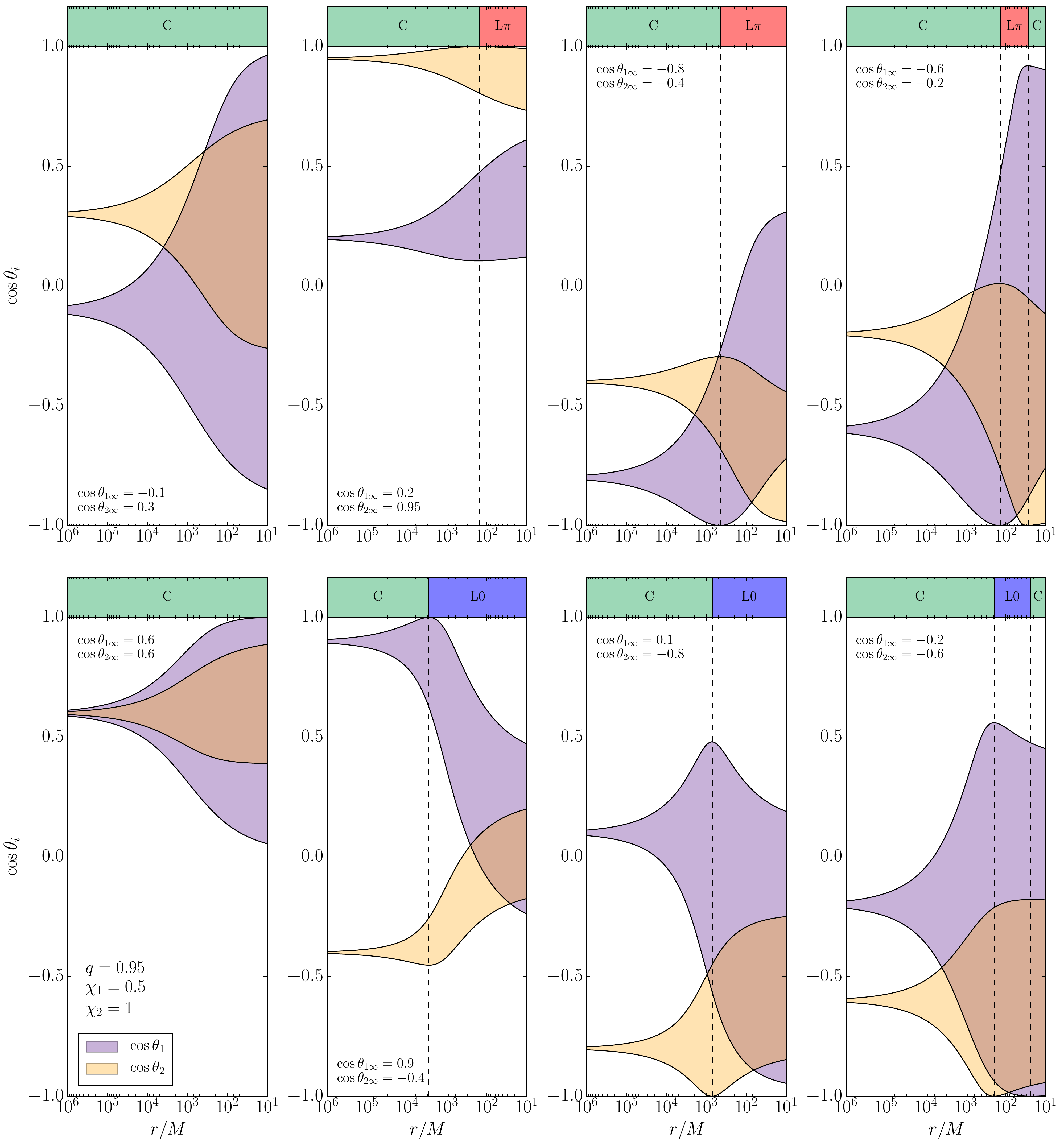}
\caption{Evolution of the spin morphology and the allowed ranges of the spin angles $\theta_i$ over a precession cycle as
functions of the binary separation $r$.  Each panel shows the range of  $\cos\theta_1$ (purple/darker) and $\cos\theta_2$
(orange/lighter) for different initial conditions $\cos\theta_{i\infty}$.  The current morphology is tracked by the horizontal
bar above each panel.  Morphologies are indicated as {\bf C} (green) for circulating, {\bf L0} (blue) for $\Delta\Phi$ librating
about 0, and {\bf L$\boldsymbol{\pi}$} (red) for $\Delta\Phi$ librating about $\pi$.  The morphology changes whenever
$\cos\theta_i=\pm 1$ (vertical dashed lines).  BBHs in the leftmost column do not undergo any transitions in the PN regime;
one transition into a librating morphology occurs for BBHs in the center columns; two transitions (circulating to librating, librating to circulating) occur for BBHs in the rightmost column.  The mass ratio and spin magnitudes are $q=0.95$,
$\chi_1=0.5$, and $\chi_2=1$ in all panels.}
\label{phasetransition}
\end{figure*}

A more complete phenomenology of phase transitions is illustrated in Fig.~\ref{phasetransition}.
The evolution of $\cos\theta_{1}$ and $\cos\theta_{2}$ along the inspiral is shown for a variety of initial conditions $\cos
\theta_{i \infty}$.  At each separation $r$, the angles $\theta_{i}$ vary on the precession time within a finite range specified
by the conditions $\xi=\xi_{\pm}(S)$ (cf. Fig.~\ref{angle_solutions}).  These envelopes vary on the radiation-reaction time as
$J$ evolves according to Eq.~(\ref{ourODE}); their width shrinks to a zero as $r/M\to\infty$ according to Eqs.~(\ref{E:csinfty}),
and tends to thicken at smaller separations because of the increasing importance of terms proportional to $S^2$ in
Eqs.~(\ref{t1t2dphisolutions}).  Horizontal bars above each panel track the binary morphologies, which we label as {\bf C},
{\bf L0}, and {\bf L$\boldsymbol{\pi}$} for circulation, libration about $\Delta\Phi = 0$, and libration about $\Delta\Phi = \pi$.
These morphologies change whenever one of the allowed ranges reach the boundaries $\cos\theta_i=\pm 1$. 

All binaries  circulate at large separation because the angles $\cos\theta_1$ and $\cos\theta_2$ are approximately
constant (Sec.~\ref{largeseplim}) and $\Delta\Phi$ from Eq.~(\ref{E:dphi}) is monotonic in $S$, thus satisfying
Eq.~(\ref{dphicirculation}).  Some binaries (leftmost panels of Fig.~\ref{phasetransition}) remain in the circulating
morphology until the PN approximation breaks down ($r=10M$).  Other binaries undergo a single transition into a librating
phase (middle columns of Fig.~\ref{phasetransition}); $\Delta\Phi$ will oscillate about 0 ($\pi$) following this transition if the
alignment condition $\sin\theta_i=0$ is satisfied at $S_-$ ($S_+$).
Since $\cos\theta_1$ ($\cos\theta_2$) decreases (increases) monotonically with $S$ [cf. Eqs.~(\ref{t1t2dphisolutions})], the
above conditions can be summarized as\begin{subequations}
\label{firsttransition}
\begin{align}
 \cos\theta_1=1 \quad{\rm or} \quad\cos\theta_2=-1\,&:
\quad {\bf C} \longrightarrow {\bf L0} \,, \\
 \cos\theta_1=-1 \quad{\rm or} \quad\cos\theta_2=1\,&:
\quad {\bf C} \longrightarrow {\bf L\boldsymbol{\pi}} \,.
\end{align}
\end{subequations}
These phase transitions were seen in previous (orbit-averaged) simulations \cite{2004PhRvD..70l4020S} and referred to
as \emph{spin locking}, because the BBH spins locked into libration about the spin-orbit resonances at $\xi_{\rm min}$ and
$\xi_{\rm max}$.  As the the librating binaries continue to inspiral, some may transition back into the circulating phase, as
pictured in the rightmost column of Fig.~\ref{phasetransition}.  The conditions for this second transition are
\begin{subequations}\label{secondtransition}
\begin{align}
\quad \cos\theta_1=-1 \quad{\rm or}\quad \cos\theta_2=1\,&:\quad {\bf L0} \longrightarrow {\bf C}\,,
\\
\quad \cos\theta_1=1 \quad{\rm or} \quad\cos\theta_2=-1\,&:\quad {\bf L\boldsymbol{\pi}} \longrightarrow {\bf C} \,.
\end{align}
\end{subequations}
As discussed further in Sec.~\ref{msasymmetry} below, this second phase transition occurs in the PN regime ($r \gtrsim
10M$) only in some corners of the parameter space ($q\lesssim 1$ and $\chi_1 \neq \chi_2$).  We have not found any
additional transitions in the PN regime, but multiple transitions may occur at the smaller separations accessible to
numerical-relativity simulations.

\begin{figure*}[p]
\includegraphics[width=\textwidth]{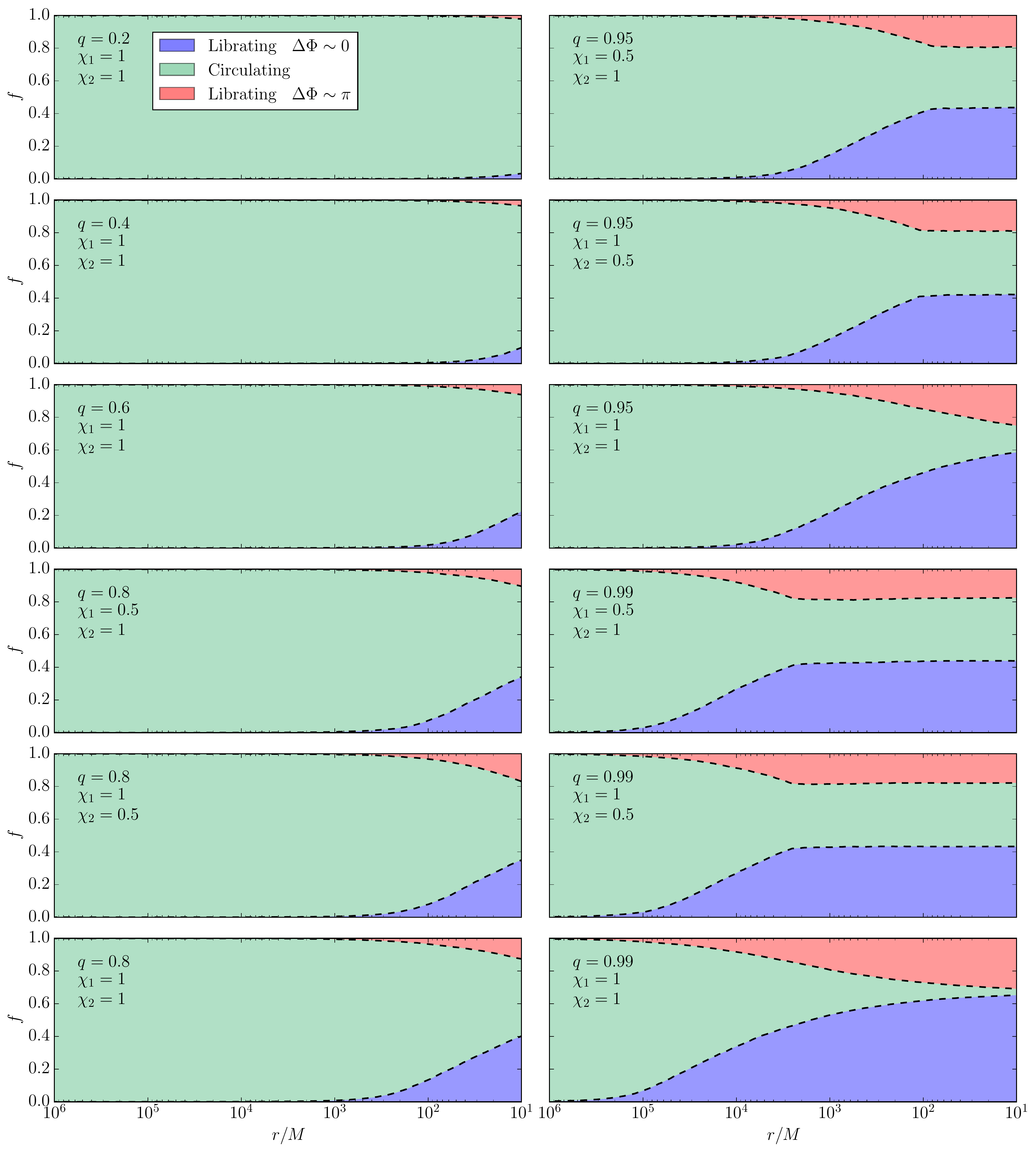}
\caption{The fraction $f$ of isotropic binaries in each of the three precessional morphologies as functions of the binary
separation. Each panel refers to different values of $q$, $\chi_1$ and $\chi_2$ as indicated in the legends. The fraction of
binaries in which $\Delta\Phi$ circulates (green, middle region of each panel), oscillates about 0 (blue, bottom region of
each panel), or oscillates about $\pi$ (red, top region of each panel) is shown as the binary orbit shrinks, with dashed lines
separating the different morphologies. The fraction of binaries in librating morphologies generally grows during the inspiral; 
this growth is stronger as $q\to 1$ but may stall for nearly equal masses and $\chi_1\neq \chi_2$, as seen in panels in the
right column.
}
\label{followtransitions}
\end{figure*}

\subsection{Dependence on mass and spin asymmetry}
\label{msasymmetry}

The asymmetry in the masses $m_i$ and spin magnitudes $\chi_i$ determines which of the eight scenarios depicted in
Fig.~\ref{phasetransition} a binary will experience during its inspiral.  The alignment conditions $\sin\theta_1=0$ and $\sin
\theta_2=0$ tend to be satisfied at similar values of $\xi$ for symmetric binaries ($q\to 1$ and $\chi_1\simeq\chi_2$),
shrinking the circulating (green) region in the left panel of Fig.~\ref{threeeffpot} and enhancing the fraction of librating
binaries.  
\begin{figure*}[p]
\centering
\includegraphics[width=\textwidth]{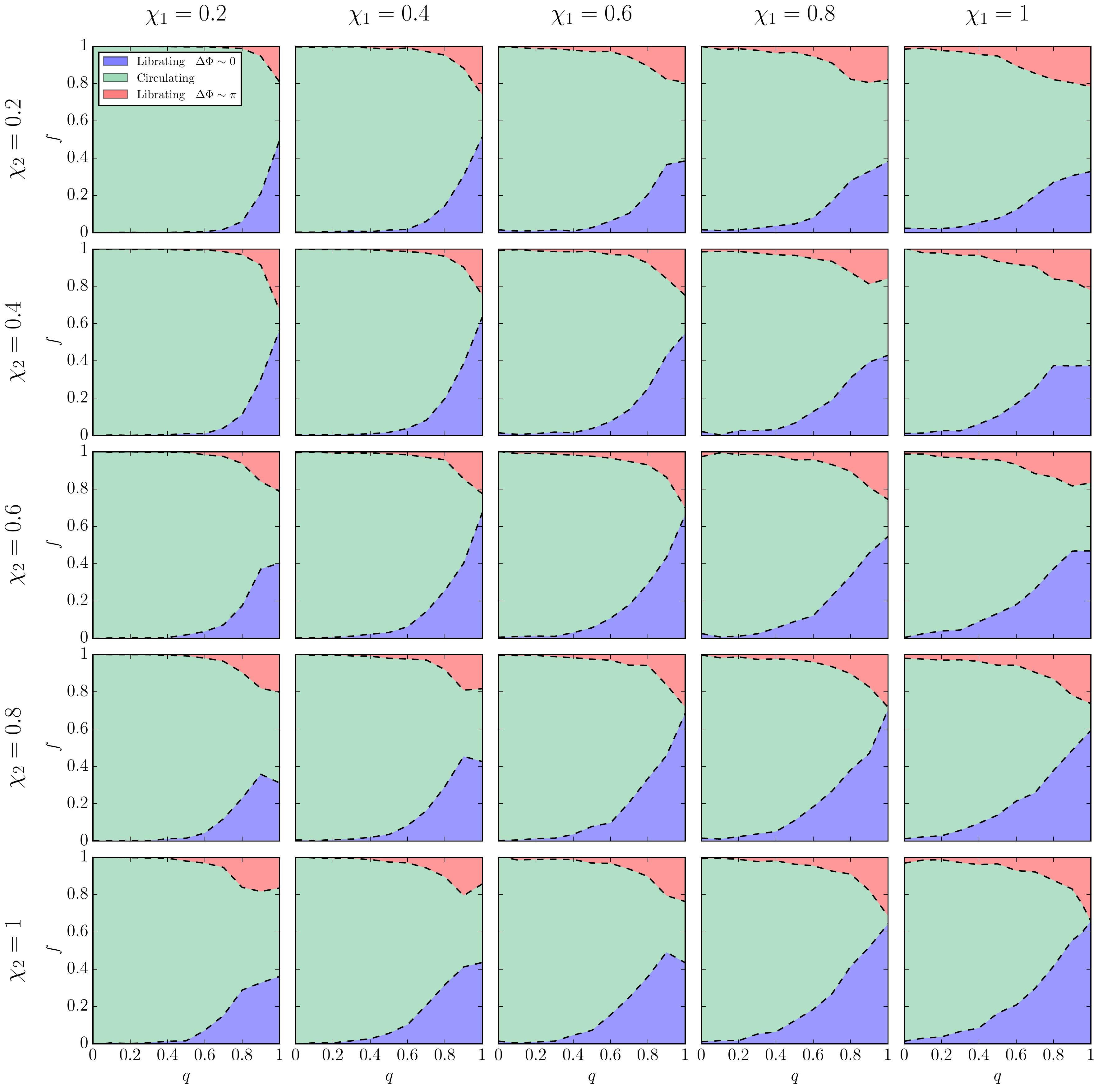}
\caption{The fraction $f$ of isotropic BBHs for which $\Delta\Phi$ circulates (green, middle region), oscillates about 0 (blue,
bottom region), or oscillates about $\pi$ (red, top region) at a binary separation $r = 10 M$ as functions of the mass ratio
$q$.  Dashed lines separate the different morphologies.  Each panel corresponds to a different value of $\chi_1$ (columns)
and $\chi_2$ (rows).  The fraction of BBHs in librating morphologies increases as the mass asymmetry decreases
($q\to 1$).  For nearly equal masses ($q\gtrsim 0.9$), asymmetry in the spin magnitudes increases the fraction of binaries
in the circulating morphology as can be seen by comparing panels on and off of the diagonal.
Some data used in this plot are listed in Table~\ref{gridtable}.  The website~\cite{DGwebsite} contains an animated version
of this figure, where the panels are shown at decreasing binary separations.}
\label{gridtransitions}
\end{figure*}
\begin{table*}[h] 
\centering
\begin{tabular}{ l|c|c|c|c|c| }
\multicolumn{1}{r}{}
&  \multicolumn{1}{c}{$\chi_1=0.2$}
&  \multicolumn{1}{c}{$\chi_1=0.4$}
&  \multicolumn{1}{c}{$\chi_1=0.6$}
&  \multicolumn{1}{c}{$\chi_1=0.8$}
&  \multicolumn{1}{c}{$\chi_1=1$}
\\[7pt]
\cline{2-6}
\rotatebox[origin=c]{90}{$\chi_2=0.2$}$\quad$
&
\begin{tabular}{@{\hskip 5pt}cccc@{\hskip 5pt}}
&&&\\[-4pt]
$q$ & \textcolor{NavyBlue}{\textbf{L0}} & \textcolor{ForestGreen}{\textbf{C}}  & \textcolor{RedOrange}{\textbf{L$\pi$}} \\
0.05 & 0.00 & 1.00 & 0.00 \\
0.2 & 0.00 & 1.00 & 0.00 \\
0.4 & 0.00 & 1.00 & 0.00 \\
0.6 & 0.01 & 0.99 & 0.00 \\
0.8 & 0.06 & 0.93 & 0.01 \\
0.95 & 0.35 & 0.53 & 0.12 \\
&&&\\[-4pt]
\end{tabular}
& 
\begin{tabular}{@{\hskip 5pt}c@{\hskip 4pt}c@{\hskip 4pt}c@{\hskip 4pt}c@{\hskip 5pt}}
&&&\\[-4pt]
$q$ & \textcolor{NavyBlue}{\textbf{L0}} & \textcolor{ForestGreen}{\textbf{C}}  & \textcolor{RedOrange}{\textbf{L$\pi$}} \\
0.05 & 0.00 & 0.99 & 0.00 \\
0.2 & 0.01 & 0.99 & 0.00 \\
0.4 & 0.01 & 0.98 & 0.01 \\
0.6 & 0.02 & 0.97 & 0.01 \\
0.8 & 0.14 & 0.81 & 0.05 \\
0.95 & 0.41 & 0.40 & 0.19 \\
&&&\\[-4pt]
\end{tabular}
&
\begin{tabular}{@{\hskip 5pt}c@{\hskip 4pt}c@{\hskip 4pt}c@{\hskip 4pt}c@{\hskip 5pt}}
&&&\\[-4pt]
$q$ & \textcolor{NavyBlue}{\textbf{L0}} & \textcolor{ForestGreen}{\textbf{C}}  & \textcolor{RedOrange}{\textbf{L$\pi$}} \\
0.05 & 0.01 & 0.98 & 0.01 \\
0.2 & 0.01 & 0.98 & 0.01 \\
0.4 & 0.01 & 0.97 & 0.02 \\
0.6 & 0.06 & 0.91 & 0.03 \\
0.8 & 0.20 & 0.69 & 0.11 \\
0.95 & 0.38 & 0.44 & 0.18 \\
&&&\\[-4pt]
\end{tabular}
&
\begin{tabular}{@{\hskip 5pt}c@{\hskip 4pt}c@{\hskip 4pt}c@{\hskip 4pt}c@{\hskip 5pt}}
&&&\\[-4pt]
$q$ & \textcolor{NavyBlue}{\textbf{L0}} & \textcolor{ForestGreen}{\textbf{C}}  & \textcolor{RedOrange}{\textbf{L$\pi$}} \\
0.05 & 0.01 & 0.98 & 0.01 \\
0.2 & 0.02 & 0.97 & 0.01 \\
0.4 & 0.04 & 0.93 & 0.04 \\
0.6 & 0.08 & 0.86 & 0.06 \\
0.8 & 0.28 & 0.54 & 0.18 \\
0.95 & 0.35 & 0.46 & 0.19 \\
&&&\\[-4pt]
\end{tabular}
&
\begin{tabular}{@{\hskip 5pt}c@{\hskip 4pt}c@{\hskip 4pt}c@{\hskip 4pt}c@{\hskip 5pt}}
&&&\\[-4pt]
$q$ & \textcolor{NavyBlue}{\textbf{L0}} & \textcolor{ForestGreen}{\textbf{C}}  & \textcolor{RedOrange}{\textbf{L$\pi$}} \\
0.05 & 0.02 & 0.96 & 0.01 \\
0.2 & 0.02 & 0.95 & 0.02 \\
0.4 & 0.06 & 0.90 & 0.04 \\
0.6 & 0.12 & 0.77 & 0.10 \\
0.8 & 0.27 & 0.55 & 0.18 \\
0.95 & 0.32 & 0.48 & 0.21 \\
&&&\\[-4pt]
\end{tabular}
\\ 
\cline{2-6}
\rotatebox[origin=c]{90}{$\chi_2=0.4$}$\quad$
&
\begin{tabular}{@{\hskip 5pt}c@{\hskip 4pt}c@{\hskip 4pt}c@{\hskip 4pt}c@{\hskip 5pt}}
&&&\\[-4pt]
$q$ & \textcolor{NavyBlue}{\textbf{L0}} & \textcolor{ForestGreen}{\textbf{C}}  & \textcolor{RedOrange}{\textbf{L$\pi$}} \\
0.05 & 0.00 & 1.00 & 0.00 \\
0.2 & 0.00 & 1.00 & 0.00 \\
0.4 & 0.00 & 0.99 & 0.00 \\
0.6 & 0.01 & 0.99 & 0.00 \\
0.8 & 0.11 & 0.86 & 0.03 \\
0.95 & 0.43 & 0.36 & 0.21 \\
&&&\\[-4pt]
\end{tabular}
& 
\begin{tabular}{@{\hskip 5pt}c@{\hskip 4pt}c@{\hskip 4pt}c@{\hskip 4pt}c@{\hskip 5pt}}
&&&\\[-4pt]
$q$ & \textcolor{NavyBlue}{\textbf{L0}} & \textcolor{ForestGreen}{\textbf{C}}  & \textcolor{RedOrange}{\textbf{L$\pi$}} \\
0.05 & 0.00 & 0.99 & 0.00 \\
0.2 & 0.00 & 0.99 & 0.00 \\
0.4 & 0.01 & 0.99 & 0.00 \\
0.6 & 0.04 & 0.95 & 0.01 \\
0.8 & 0.20 & 0.76 & 0.04 \\
0.95 & 0.51 & 0.32 & 0.17 \\
&&&\\[-4pt]
\end{tabular}
&
\begin{tabular}{@{\hskip 5pt}c@{\hskip 4pt}c@{\hskip 4pt}c@{\hskip 4pt}c@{\hskip 5pt}}
&&&\\[-4pt]
$q$ & \textcolor{NavyBlue}{\textbf{L0}} & \textcolor{ForestGreen}{\textbf{C}}  & \textcolor{RedOrange}{\textbf{L$\pi$}} \\
0.05 & 0.01 & 0.98 & 0.01 \\
0.2 & 0.01 & 0.98 & 0.01 \\
0.4 & 0.01 & 0.97 & 0.01 \\
0.6 & 0.07 & 0.90 & 0.03 \\
0.8 & 0.25 & 0.67 & 0.08 \\
0.95 & 0.49 & 0.31 & 0.20 \\
&&&\\[-4pt]
\end{tabular}
&
\begin{tabular}{@{\hskip 5pt}c@{\hskip 4pt}c@{\hskip 4pt}c@{\hskip 4pt}c@{\hskip 5pt}}
&&&\\[-4pt]
$q$ & \textcolor{NavyBlue}{\textbf{L0}} & \textcolor{ForestGreen}{\textbf{C}}  & \textcolor{RedOrange}{\textbf{L$\pi$}} \\
0.05 & 0.01 & 0.97 & 0.01 \\
0.2 & 0.03 & 0.96 & 0.01 \\
0.4 & 0.03 & 0.94 & 0.03 \\
0.6 & 0.13 & 0.82 & 0.05 \\
0.8 & 0.31 & 0.57 & 0.13 \\
0.95 & 0.41 & 0.41 & 0.17 \\
&&&\\[-4pt]
\end{tabular}
&
\begin{tabular}{@{\hskip 5pt}c@{\hskip 4pt}c@{\hskip 4pt}c@{\hskip 4pt}c@{\hskip 5pt}}
&&&\\[-4pt]
$q$ & \textcolor{NavyBlue}{\textbf{L0}} & \textcolor{ForestGreen}{\textbf{C}}  & \textcolor{RedOrange}{\textbf{L$\pi$}} \\
0.05 & 0.01 & 0.98 & 0.01 \\
0.2 & 0.03 & 0.95 & 0.02 \\
0.4 & 0.06 & 0.91 & 0.03 \\
0.6 & 0.17 & 0.75 & 0.08 \\
0.8 & 0.38 & 0.46 & 0.16 \\
0.95 & 0.37 & 0.43 & 0.20 \\
&&&\\[-4pt]
\end{tabular}
\\
\cline{2-6}
\rotatebox[origin=c]{90}{$\chi_2=0.6$}$\quad$
&
\begin{tabular}{@{\hskip 5pt}c@{\hskip 4pt}c@{\hskip 4pt}c@{\hskip 4pt}c@{\hskip 5pt}}
&&&\\[-4pt]
$q$ & \textcolor{NavyBlue}{\textbf{L0}} & \textcolor{ForestGreen}{\textbf{C}}  & \textcolor{RedOrange}{\textbf{L$\pi$}} \\
0.05 & 0.00 & 1.00 & 0.00 \\
0.2 & 0.00 & 1.00 & 0.00 \\
0.4 & 0.00 & 0.99 & 0.00 \\
0.6 & 0.04 & 0.95 & 0.02 \\
0.8 & 0.17 & 0.76 & 0.06 \\
0.95 & 0.39 & 0.43 & 0.19 \\
&&&\\[-4pt]
\end{tabular}
& 
\begin{tabular}{@{\hskip 5pt}c@{\hskip 4pt}c@{\hskip 4pt}c@{\hskip 4pt}c@{\hskip 5pt}}
&&&\\[-4pt]
$q$ & \textcolor{NavyBlue}{\textbf{L0}} & \textcolor{ForestGreen}{\textbf{C}}  & \textcolor{RedOrange}{\textbf{L$\pi$}} \\
0.05 & 0.00 & 0.99 & 0.00 \\
0.2 & 0.01 & 0.99 & 0.01 \\
0.4 & 0.02 & 0.97 & 0.01 \\
0.6 & 0.06 & 0.92 & 0.02 \\
0.8 & 0.26 & 0.70 & 0.04 \\
0.95 & 0.54 & 0.28 & 0.18 \\
&&&\\[-4pt]
\end{tabular}
&
\begin{tabular}{@{\hskip 5pt}c@{\hskip 4pt}c@{\hskip 4pt}c@{\hskip 4pt}c@{\hskip 5pt}}
&&&\\[-4pt]
$q$ & \textcolor{NavyBlue}{\textbf{L0}} & \textcolor{ForestGreen}{\textbf{C}}  & \textcolor{RedOrange}{\textbf{L$\pi$}} \\
0.05 & 0.01 & 0.99 & 0.00 \\
0.2 & 0.01 & 0.98 & 0.01 \\
0.4 & 0.03 & 0.95 & 0.02 \\
0.6 & 0.11 & 0.86 & 0.03 \\
0.8 & 0.29 & 0.64 & 0.07 \\
0.95 & 0.55 & 0.23 & 0.22 \\
&&&\\[-4pt]
\end{tabular}
&
\begin{tabular}{@{\hskip 5pt}c@{\hskip 4pt}c@{\hskip 4pt}c@{\hskip 4pt}c@{\hskip 5pt}}
&&&\\[-4pt]
$q$ & \textcolor{NavyBlue}{\textbf{L0}} & \textcolor{ForestGreen}{\textbf{C}}  & \textcolor{RedOrange}{\textbf{L$\pi$}} \\
0.05 & 0.02 & 0.97 & 0.02 \\
0.2 & 0.01 & 0.97 & 0.01 \\
0.4 & 0.06 & 0.92 & 0.02 \\
0.6 & 0.12 & 0.84 & 0.04 \\
0.8 & 0.33 & 0.56 & 0.10 \\
0.95 & 0.50 & 0.28 & 0.22 \\
&&&\\[-4pt]
\end{tabular}
&
\begin{tabular}{@{\hskip 5pt}c@{\hskip 4pt}c@{\hskip 4pt}c@{\hskip 4pt}c@{\hskip 5pt}}
&&&\\[-4pt]
$q$ & \textcolor{NavyBlue}{\textbf{L0}} & \textcolor{ForestGreen}{\textbf{C}}  & \textcolor{RedOrange}{\textbf{L$\pi$}} \\
0.05 & 0.01 & 0.97 & 0.01 \\
0.2 & 0.04 & 0.93 & 0.03 \\
0.4 & 0.09 & 0.87 & 0.04 \\
0.6 & 0.18 & 0.75 & 0.07 \\
0.8 & 0.37 & 0.49 & 0.14 \\
0.95 & 0.47 & 0.36 & 0.17 \\
&&&\\[-4pt]
\end{tabular}
\\
\cline{2-6}
\rotatebox[origin=c]{90}{$\chi_2=0.8$}$\quad$
&
\begin{tabular}{@{\hskip 5pt}c@{\hskip 4pt}c@{\hskip 4pt}c@{\hskip 4pt}c@{\hskip 5pt}}
&&&\\[-4pt]
$q$ & \textcolor{NavyBlue}{\textbf{L0}} & \textcolor{ForestGreen}{\textbf{C}}  & \textcolor{RedOrange}{\textbf{L$\pi$}} \\
0.05 & 0.00 & 1.00 & 0.00 \\
0.2 & 0.00 & 1.00 & 0.00 \\
0.4 & 0.01 & 0.98 & 0.01 \\
0.6 & 0.04 & 0.94 & 0.02 \\
0.8 & 0.23 & 0.68 & 0.10 \\
0.95 & 0.34 & 0.47 & 0.19 \\
&&&\\[-4pt]
\end{tabular}
& 
\begin{tabular}{@{\hskip 5pt}c@{\hskip 4pt}c@{\hskip 4pt}c@{\hskip 4pt}c@{\hskip 5pt}}
&&&\\[-4pt]
$q$ & \textcolor{NavyBlue}{\textbf{L0}} & \textcolor{ForestGreen}{\textbf{C}}  & \textcolor{RedOrange}{\textbf{L$\pi$}} \\
0.05 & 0.00 & 0.99 & 0.00 \\
0.2 & 0.01 & 0.99 & 0.00 \\
0.4 & 0.02 & 0.97 & 0.01 \\
0.6 & 0.08 & 0.89 & 0.02 \\
0.8 & 0.29 & 0.63 & 0.08 \\
0.95 & 0.44 & 0.37 & 0.19 \\
&&&\\[-4pt]
\end{tabular}
&
\begin{tabular}{@{\hskip 5pt}c@{\hskip 4pt}c@{\hskip 4pt}c@{\hskip 4pt}c@{\hskip 5pt}}
&&&\\[-4pt]
$q$ & \textcolor{NavyBlue}{\textbf{L0}} & \textcolor{ForestGreen}{\textbf{C}}  & \textcolor{RedOrange}{\textbf{L$\pi$}} \\
0.05 & 0.00 & 0.99 & 0.01 \\
0.2 & 0.01 & 0.98 & 0.01 \\
0.4 & 0.04 & 0.95 & 0.02 \\
0.6 & 0.10 & 0.87 & 0.03 \\
0.8 & 0.34 & 0.61 & 0.06 \\
0.95 & 0.57 & 0.21 & 0.22 \\
&&&\\[-4pt]
\end{tabular}
&
\begin{tabular}{@{\hskip 5pt}c@{\hskip 4pt}c@{\hskip 4pt}c@{\hskip 4pt}c@{\hskip 5pt}}
&&&\\[-4pt]
$q$ & \textcolor{NavyBlue}{\textbf{L0}} & \textcolor{ForestGreen}{\textbf{C}}  & \textcolor{RedOrange}{\textbf{L$\pi$}} \\
0.05 & 0.01 & 0.98 & 0.01 \\
0.2 & 0.02 & 0.96 & 0.01 \\
0.4 & 0.05 & 0.93 & 0.02 \\
0.6 & 0.18 & 0.78 & 0.04 \\
0.8 & 0.38 & 0.52 & 0.10 \\
0.95 & 0.59 & 0.18 & 0.23 \\
&&&\\[-4pt]
\end{tabular}
&
\begin{tabular}{@{\hskip 5pt}c@{\hskip 4pt}c@{\hskip 4pt}c@{\hskip 4pt}c@{\hskip 5pt}}
&&&\\[-4pt]
$q$ & \textcolor{NavyBlue}{\textbf{L0}} & \textcolor{ForestGreen}{\textbf{C}}  & \textcolor{RedOrange}{\textbf{L$\pi$}} \\
0.05 & 0.02 & 0.96 & 0.02 \\
0.2 & 0.03 & 0.94 & 0.03 \\
0.4 & 0.10 & 0.87 & 0.04 \\
0.6 & 0.21 & 0.73 & 0.06 \\
0.8 & 0.38 & 0.49 & 0.13 \\
0.95 & 0.54 & 0.22 & 0.24 \\
&&&\\[-4pt]
\end{tabular}
\\
\cline{2-6}
\rotatebox[origin=c]{90}{$\chi_2=1$}$\quad$
&
\begin{tabular}{@{\hskip 5pt}c@{\hskip 4pt}c@{\hskip 4pt}c@{\hskip 4pt}c@{\hskip 5pt}}
&&&\\[-4pt]
$q$ & \textcolor{NavyBlue}{\textbf{L0}} & \textcolor{ForestGreen}{\textbf{C}}  & \textcolor{RedOrange}{\textbf{L$\pi$}} \\
0.05 & 0.00 & 1.00 & 0.00 \\
0.2 & 0.00 & 1.00 & 0.00 \\
0.4 & 0.01 & 0.98 & 0.01 \\
0.6 & 0.07 & 0.90 & 0.03 \\
0.8 & 0.29 & 0.55 & 0.16 \\
0.95 & 0.34 & 0.48 & 0.17 \\
&&&\\[-4pt]
\end{tabular}
& 
\begin{tabular}{@{\hskip 5pt}c@{\hskip 4pt}c@{\hskip 4pt}c@{\hskip 4pt}c@{\hskip 5pt}}
&&&\\[-4pt]
$q$ & \textcolor{NavyBlue}{\textbf{L0}} & \textcolor{ForestGreen}{\textbf{C}}  & \textcolor{RedOrange}{\textbf{L$\pi$}} \\
0.05 & 0.00 & 1.00 & 0.00 \\
0.2 & 0.01 & 0.99 & 0.01 \\
0.4 & 0.03 & 0.96 & 0.01 \\
0.6 & 0.10 & 0.87 & 0.03 \\
0.8 & 0.32 & 0.58 & 0.10 \\
0.95 & 0.42 & 0.40 & 0.17 \\
&&&\\[-4pt]
\end{tabular}
&
\begin{tabular}{@{\hskip 5pt}c@{\hskip 4pt}c@{\hskip 4pt}c@{\hskip 4pt}c@{\hskip 5pt}}
&&&\\[-4pt]
$q$ & \textcolor{NavyBlue}{\textbf{L0}} & \textcolor{ForestGreen}{\textbf{C}}  & \textcolor{RedOrange}{\textbf{L$\pi$}} \\
0.05 & 0.01 & 0.99 & 0.00 \\
0.2 & 0.01 & 0.98 & 0.01 \\
0.4 & 0.05 & 0.94 & 0.01 \\
0.6 & 0.15 & 0.81 & 0.03 \\
0.8 & 0.36 & 0.54 & 0.10 \\
0.95 & 0.46 & 0.32 & 0.22 \\
&&&\\[-4pt]
\end{tabular}
&
\begin{tabular}{@{\hskip 5pt}c@{\hskip 4pt}c@{\hskip 4pt}c@{\hskip 4pt}c@{\hskip 5pt}}
&&&\\[-4pt]
$q$ & \textcolor{NavyBlue}{\textbf{L0}} & \textcolor{ForestGreen}{\textbf{C}}  & \textcolor{RedOrange}{\textbf{L$\pi$}} \\
0.05 & 0.01 & 0.98 & 0.01 \\
0.2 & 0.02 & 0.97 & 0.01 \\
0.4 & 0.06 & 0.92 & 0.02 \\
0.6 & 0.18 & 0.77 & 0.04 \\
0.8 & 0.41 & 0.49 & 0.09 \\
0.95 & 0.58 & 0.17 & 0.25 \\
&&&\\[-4pt]
\end{tabular}
&
\begin{tabular}{@{\hskip 5pt}c@{\hskip 4pt}c@{\hskip 4pt}c@{\hskip 4pt}c@{\hskip 5pt}}
&&&\\[-4pt]
$q$ & \textcolor{NavyBlue}{\textbf{L0}} & \textcolor{ForestGreen}{\textbf{C}}  & \textcolor{RedOrange}{\textbf{L$\pi$}} \\
0.05 & 0.02 & 0.95 & 0.02 \\
0.2 & 0.04 & 0.95 & 0.01 \\
0.4 & 0.08 & 0.88 & 0.04 \\
0.6 & 0.21 & 0.72 & 0.07 \\
0.8 & 0.42 & 0.46 & 0.12 \\
0.95 & 0.59 & 0.16 & 0.24 \\
&&&\\[-4pt]
\end{tabular}
\\
\cline{2-6}
\end{tabular}

\caption{Fractions of isotropic BBHs in each of the three precessional morphologies (\textcolor{NavyBlue}{\textbf{L0}}:
$\Delta\Phi$ oscillates about 0, \textcolor{ForestGreen}{\textbf{C}}: $\Delta\Phi$ circulates,
\textcolor{RedOrange}{\textbf{L$\pi$}}: $\Delta\Phi$ oscillates about $\pi$) at $r=10M$ as shown in Fig.~\ref{gridtransitions}. 
For a grid of values in $\chi_1$ (columns), $\chi_2$ (rows) and, $q$ (first column in each mini-table), we report the fraction
of binaries in each morphology.  The sum of the three fractions may differ from unity because of rounding errors.
}
\label{gridtable}
\end{table*}
This point is illustrated in Fig.~\ref{followtransitions} below, which shows the fraction of isotropic binaries in each of the three
morphologies as functions of the binary separation.  Each panel is computed by averaging over  a sample of binaries
isotropically distributed at large separations (flat distributions in $\cos\theta_{1\infty}$ and $\cos\theta_{2\infty}$); all
binaries in each sample share the same mass ratio and spin magnitudes.  As the separation decreases, binaries transition
from the circulating to librating morphologies.  The fraction of binaries experiencing these transitions strongly depends on 
the  mass ratio $q$. If the mass ratio is low ($q\lesssim 0.6$), most binaries remain circulating down to very small
separations $r\sim 10M$.  Comparable-mass binaries ($q\gtrsim 0.6$) are more likely to undergo a phase transition in the
PN regime.  The typical transition radius $r_{\rm tr}$ at which these phase transitions occur is also very sensitive to the
mass ratio \cite{2004PhRvD..70l4020S,2010PhRvD..81h4054K}; transitions occur in the very late inspiral for low mass ratios while $r_{\rm tr}$ can
be as large as $10^5M$ for $q\simeq0.99$.  Very long evolutions are needed to capture all of the morphological transitions
for nearly equal-mass binaries; such long inspirals are prohibitively expensive in the standard orbit-averaged approach (as
seen in Fig.~\ref{timing}) but can easily be calculated within our new precession-averaged formalism.

A more extensive exploration of how BBH spin morphology depends on the binary parameters is shown in
Fig.~\ref{gridtransitions} and Table~\ref{gridtable}.  Isotropic distributions at $r/M=\infty$ are evolved down to $r=10M$,
where their morphologies are determined; as shown in the upper panel of Fig.~\ref{benchmark}, these initially isotropic
distributions remain isotropic at smaller separations.
The fraction of binaries in each morphology at $r=10M$ is shown as functions of $q$ for a grid of values of the spin
magnitudes $\chi_1$ and $\chi_2$.  As was already seen in Fig.~\ref{followtransitions}, the likelihood of phase transitions
depends on the mass ratio $q$; more librating binaries are found for comparable-mass BBHs at any fixed separation.

Spin magnitudes also affect the fraction of BBHs in each morphology.  As one moves along the diagonal of
Fig.~\ref{gridtransitions} in the direction of increasing $\chi_1 = \chi_2$, a slightly higher fraction of binaries are found in 
librating morphologies because of increased spin-spin coupling \cite{2004PhRvD..70l4020S}.  The corner of the parameter space characterized by mass
symmetry and spin asymmetry ($q\to 1$ and $\chi_1\neq \chi_2$) presents a peculiar phenomenology, as seen in the right
panels of Fig.~\ref{followtransitions}, where the fraction of binaries in each morphology approaches constant values for
$r \lesssim 1000M$.  This behavior can be explained by recognizing that in this region of
parameter space binaries may undergo two morphological transitions in the PN regime, as seen in the rightmost panels of
Fig.~\ref{phasetransition}.  The number of binaries experiencing their first phase transition from circulation to libration is
nearly canceled by the number of binaries undergoing a second phase transition back to the circulating morphology,
leading to almost constant fractions of binaries in each morphology.  This effect also accounts for the kinks in the
morphology fractions at $q\simeq 0.9$ in the off-diagonal ($\chi_1 \neq \chi_2$) panels of Fig.~\ref{gridtransitions}. 

\subsection{Predicting spin morphology at small separations}
\label{binmemory}

\begin{figure*}[t]
\includegraphics[width=\textwidth]{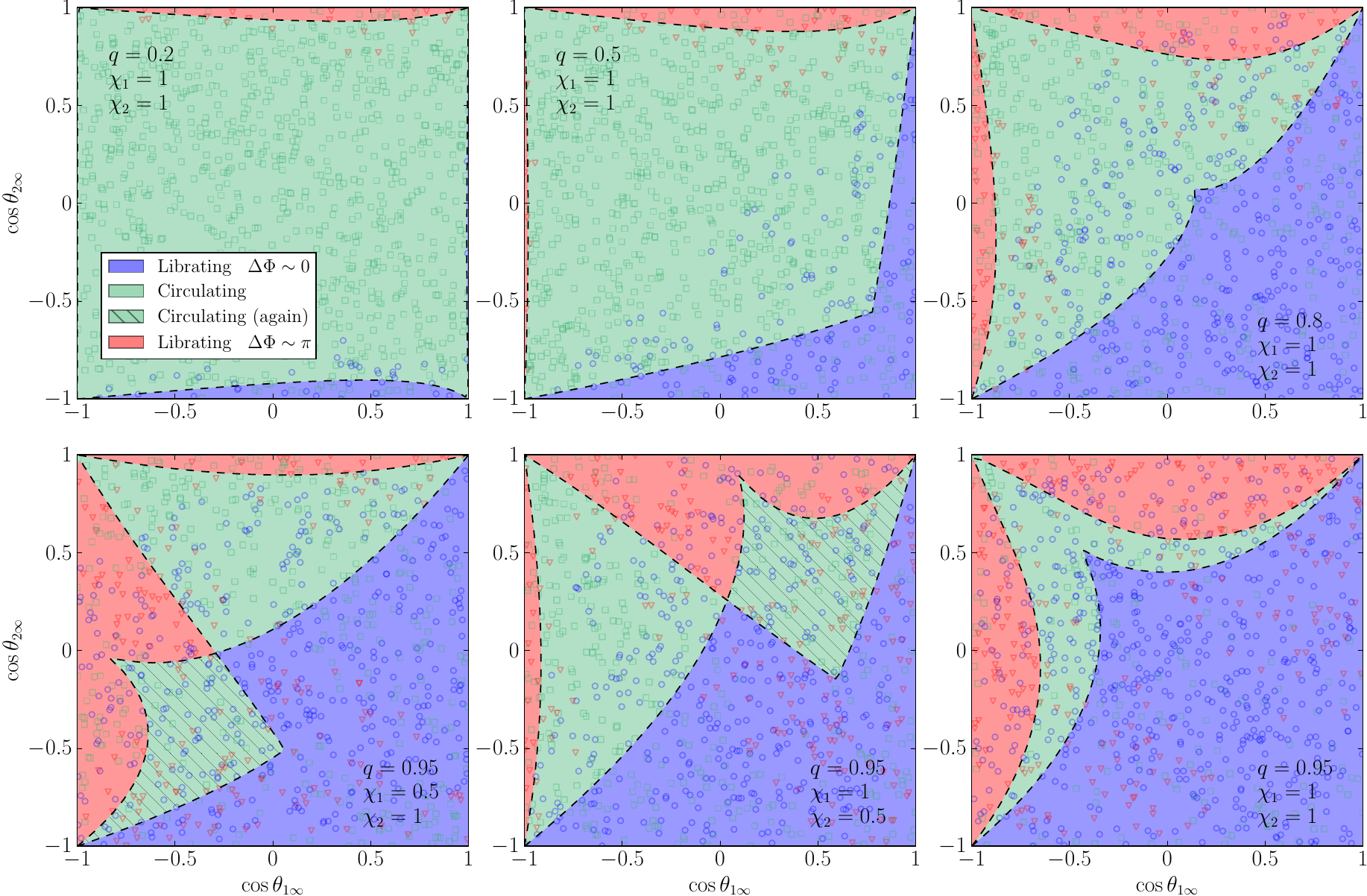}
\caption{Spin morphologies at $r_f=10M$ as functions of the asymptotic values of the spin angles $\theta_{i\infty}$.  The
mass ratio $q$ and spin magnitudes $\chi_i$ for each panel are indicated in the legends.  Evolving BBHs along the four
lines $\cos\theta_i=\pm1$ at $r_f$ out to $r/M\to\infty$ using our new precession-averaged approach yields the dashed
curves separating the different final morphologies: $\Delta\Phi$ oscillates about 0 (blue), oscillates about $\pi$ (red),
circulates without ever having experienced a phase transition (plain green), or circulates after having experienced a phase
transition to libration and then a second phase transition back to circulation (hatched green).  The morphology within
each region defined by the dashed boundaries is determined by which of the conditions $\cos\theta_i=\pm1$ these
boundaries satisfy, as described in Sec.~\ref{binmemory}.  The points show the locations of binaries in the
$\cos\theta_{1}-\cos\theta_{2}$ 
plane at $r_f$ and are colored by their morphology at that separation [$\Delta\Phi$
oscillates about 0 (blue circles), oscillates about $\pi$ (green squares), or circulates (red triangles)].  Because morphology
depends on $\Delta\Phi$ in addition to $\theta_1$ and $\theta_2$ at finite separation, the projection onto the
$\cos\theta_1-\cos\theta_2$ plane can lead points of different morphologies to occur at the same positions,
particularly for comparable-mass binaries $q\simeq1$ where the $\theta_i$'s oscillate with greater amplitude.  The
website~\cite{DGwebsite} contains an animated version of this figure in which $r_f$ evolves.}
\label{borders}
\end{figure*}

We described in great detail in Sec.~\ref{subsecmorph} how to determine the BBH spin morphology from the binary
parameters at a given separation, but astrophysical BBHs are often formed at much larger separations than where we are
interested in observing them.  Although BBHs can be efficiently evolved to smaller separations using the
precession-averaged approach described in Sec.~\ref{transferbin}, we can in fact predict the spin morphology at a final
separation $r_f$ based solely upon the asymptotic values of $\theta_{1\infty}$ and $\theta_{2\infty }$ [or equivalently $\xi$
and $\kappa_{\infty}$ according to Eqs.~(\ref{E:csinfty})] without the need to integrate $dJ/dL$ down to $r_f$.  This can be
achieved by recognizing that the curves in the $\cos\theta_{1\infty}-\cos\theta_{2\infty}$ plane separating the final
morphologies at $r_f$ correspond to BBHs experiencing phase transitions at $r_f$, i.e. binaries for which $\cos\theta_i(r_f)
= \pm1$.  These binaries constitute the four borders of the $\cos\theta_1-\cos\theta_2$ plane at $r_f$; using our
expression for $dJ/dL$ in Eq.~(\ref{ourODE}) to integrate BBHs along these borders out to $r/M \to \infty$, we obtain four
curves in the $\cos\theta_{1\infty}-\cos\theta_{2\infty}$ plane, as seen in Fig.~\ref{borders}.  These curves define regions {\bf I}
and {\bf II} in the $\cos\theta_{1\infty}-\cos\theta_{2\infty}$ plane with the following boundaries:
\begin{align*}
{\bf I.}\qquad
&\cos\theta_{1\infty} = +1,& &\cos\theta_{2\infty} = -1, \\ 
&\cos\theta_1(r_f) = +1,& &\cos\theta_2(r_f) = -1;
\\
{\bf II.}\qquad
&\cos\theta_{1\infty} = -1,  &&\cos\theta_{2\infty} = +1, \\ 
&\cos\theta_1(r_f) = -1, &&\cos\theta_2(r_f) = +1.
\end{align*}

The final morphology at $r_f$ for each point in the $\cos\theta_{1\infty}-\cos\theta_{2\infty}$ plane is determined by whether
or not that point is contained in the two regions:
\begin{itemize}

\item Outside both region {\bf I} and region {\bf II}: $\Delta\Phi$ circulates (no phase transitions, plain green  in  Fig.~\ref{borders}).

\item Inside region {\bf I} but not region {\bf II}: $\Delta\Phi$ oscillates about 0 (one phase transition,  blue  in  Fig.~\ref{borders}).

\item Inside region {\bf II} but not region {\bf I}: $\Delta\Phi$ oscillates about $\pi$ (one phase transition, red  in  Fig.~\ref{borders}).

\item Inside both region {\bf I} and region {\bf II}: $\Delta\Phi$ circulates (two phase transitions,  hatched green  in  Fig.~\ref{borders}).

\end{itemize}
These conditions on the final morphology are consistent with the criteria for phase transitions given in
Eqs.~(\ref{firsttransition}) and (\ref{secondtransition}).  Once the boundaries of regions {\bf I} and {\bf II} have been established we
can determine the final morphology of {\it any} BBH from its initial conditions at astrophysically large separations without
further need to integrate $dJ/dL$ down to $r_f$.
A binary with spin orientations lying in the green, red or blue region of Fig.~\ref{borders} at large separations will be found with $\Delta\Phi$ circulating, oscillating about $0$ or oscillating about $\pi$ at the end of the inspiral.

Measuring BBH spin morphology directly 
offers several advantages
over explicitly measuring the spin angles $\theta_1$, $\theta_2$ and $\Delta\Phi$.  Spin morphology encodes information
about BBH spin precession but is more robust than the spin angles in that it only varies on the radiation-reaction time
(being a function of $L$, $J$, and $\xi$).  Measurement of only the two angles $\theta_1$ and $\theta_2$ at small
separations constrains neither the morphology at small separations nor the initial conditions at large separations, as can be
seen from the scatter points in Fig.~\ref{borders}, which show an isotropic sample of binaries at $r_f$. Points corresponding to the circulating and both librating morphologies
lie right on top of each other in this plot, evidence of both the importance of the third angle $\Delta\Phi$ and the large
oscillations in $\theta_i$ at small separations seen in Fig.~\ref{phasetransition}.  By contrast, spin morphology is a direct
memory of a BBH's initial position in the $\cos\theta_{1\infty}-\cos\theta_{2\infty}$ plane, as seen in Fig.~\ref{borders}.
Astrophysical scenarios of BBH formation can favor some regions in this plane over others \cite{2013PhRvD..87j4028G},
implying that GW observations of spin morphology can constrain BBH formation \cite{2014PhRvD..89l4025G}.

\section{Discussion}
\label{sec:concl}

BBHs evolve on three distinct timescales: the orbital time $t_{\rm orb}$, the precession time $t_{\rm pre}$, and the
radiation-reaction time $t_{\rm RR}$.  In the PN regime ($r \gg r_g$), these timescales obey a strict hierarchy:
$t_{\rm orb} \ll t_{\rm pre} \ll t_{\rm RR}$.  All of the parameters needed to describe BBHs evolve on a distinct timescale: the
vectorial binary separation $\mathbf{r}$ on $t_{\rm orb}$, the angular-momentum directions $\hat{\mathbf{L}}$ and
$\hat{\mathbf{S}}_i$ on $t_{\rm pre}$, and the orbital-angular-momentum magnitude $L$ and total angular momentum
$\mathbf{J}$ on $t_{\rm RR}$.  The mass ratio $q$ and spin magnitudes $S_i$ remain constant throughout the inspiral.
Expanding on our previous {\it Letter} \cite{2015PhRvL.114h1103K}, we exploit this timescale hierarchy and conservation
of the projected effective spin $\xi$ \cite{2001PhRvD..64l4013D,2008PhRvD..78d4021R} throughout the inspiral to solve
the orbit-averaged 2PN equations of BBH spin precession given by Eq.~(\ref{preceq}).  The solutions given by 
Eq.~(\ref{t1t2dphisolutions}) for the three angles $\theta_1$, $\theta_2$, and $\Delta\Phi$ that specify the relative
orientations of $\mathbf{L}$, $\mathbf{S}_1$, and $\mathbf{S}_2$ are remarkably simple and are given parametrically in
terms of a single variable, the total-spin magnitude $S$, that evolves on $t_{\rm pre}$.

These solutions fully determine how the relative orientations of the three angular momenta evolve over a precession
cycle as $S$ oscillates back and forth between extrema $S_\pm$.  We find that spin precession can be classified into three
distinct morphologies depending on whether $\Delta\Phi$ oscillates about 0, oscillates about $\pi$, or circulates through
the full range $[-\pi, +\pi]$ over a precession cycle.  For BBHs with a given mass ratio $q$ and spin magnitudes $S_i$, the
precessional morphology at a binary separation $r$ is determined by $J$ and $\xi$, implying that the morphology only
evolves on the radiation-reaction time $t_{\rm RR}$.  Spin-orbit coupling dominates over the higher-PN-order spin-spin
coupling at large separations implying that all BBHs formed at such large separations begin in the circulating morphology.
Since $\xi$ is constant to high accuracy throughout the inspiral, evolving our solutions~(\ref{t1t2dphisolutions}) and their
associated morphology to smaller separations (lower values of $L$) only requires an expression for $dJ/dL$ due to
radiation reaction.  All previous studies of radiation reaction have relied on orbit-averaged expressions for
$d\mathbf{L}_{\rm RR}/dt$ that must be integrated numerically with time steps $\Delta t \lesssim t_{\rm pre}$.  Our new
solutions~(\ref{t1t2dphisolutions}) allow us to {\it precession average} these expressions to derive Eq.~(\ref{ourODE}) for
$dJ/dL$ that can be integrated with a time step $t_{\rm pre} \ll \Delta t' \lesssim t_{\rm RR}$.  The computational cost of
calculating inspirals from an initial separation $r_i$ in our new precession-averaged approach scales as $\log r_i$, leading
to vast savings over the traditionally orbit-averaged approach (which scales as $r_i^{3/2}$) for the large initial separations
relevant to astrophysical BBH formation.

Using our new expression for $dJ/dL$, we can evolve our initially circulating BBHs to smaller separations, where they may
experience a phase transition to one of the two librating morphologies.  Some of these librating BBHs may subsequently
undergo a second phase transition back to circulation before reaching a binary separation $r = 10 M$ below which the PN
approximation itself begins to break down.  Our precession-averaged calculation of the inspiral agrees well with the
orbit-averaged approach down to nearly this separation where small discrepancies appear because of dynamically
generated inhomogeneity in the precessional phase as the timescale hierarchy fails.  Unlike the angles $\theta_1$,
$\theta_2$ and $\Delta\Phi$, that vary rapidly on the precession time at small separations, the precession morphology at
small separations is directly determined by the asymptotic values $\theta_{i\infty}$ of these angles at large separations,
providing a memory of BBH formation potentially accessible to GW detectors.

Although this work focuses on BBH spin precession, our analysis also facilitates the calculation and interpretation of GW
signals.  Fast templates suitable for GW detection and parameter estimation are being developed using our new
precessional solutions and precession-averaged equation for radiation reaction~\cite{chatziioannouprep}.  The insights
underpinning our approach (most notably the use of a hierarchical coordinate system that better respects the separation of
timescales intrinsic in the binary dynamics) are also helping us to assess whether the precessional morphology of BBHs
in spin-orbit resonances can be reliably identified in the context of full GW parameter estimation~\cite{trifiroprep}.
Preliminary results indicate that BBH spin orientations can be significantly constrained at realistic signal-to-noise ratios,
suggesting that observations of BBH spin precession as described in this work may soon provide a new window into the
astrophysical origins of BBHs and general relativity itself.

\section*{Acknowledgments}

D.G. is supported by the UK STFC and the Isaac Newton Studentship of
the University of Cambridge.  M.K. is supported by Alfred P. Sloan
Foundation grant FG-2015-65299.  R.O'S. is supported by NSF grants
PHY-0970074 and PHY-1307429.  E.B.~is supported by NSF CAREER Grant
PHY-1055103 and by FCT contract IF/00797/2014/CP1214/CT0012 under the
IF2014 Programme. U.S.~is supported by FP7-PEOPLE-2011-CIG Grant
No. 293412, FP7-PEOPLE-2011-IRSES Grant No.295189, H2020 ERC
Consolidator Grant Agreement No. MaGRaTh-646597, SDSC and TACC through
XSEDE Grant No.~PHY-090003 by the NSF, Finis Terrae through Grant
No.~ICTS-CESGA-249, STFC Roller Grant No. ST/L000636/1 and DiRAC's
Cosmos Shared Memory system through BIS Grant No.~ST/J005673/1 and
STFC Grant Nos.~ST/H008586/1, ST/K00333X/1.  Figures were generated
using the \textsc{Python}-based \textsc{matplotlib} package
\citep{2007CSE.....9...90H}.

\bibliographystyle{apsrev.bst}
\bibliography{PRD}

\end{document}